\documentclass[11t, a4paper]{article}
\pdfoutput=1
\usepackage{jcappub}
\usepackage[utf8]{inputenc}
\usepackage[english]{babel}
\usepackage{amsmath, amssymb}
\usepackage{graphicx}
\usepackage{xcolor}
\usepackage{hyperref}

\makeatletter
\newcommand*{\dd}{\mathop{}\!{\operator@font d}}
\makeatother

\newcommand*{\bk}{\mathbf{k}}
\newcommand*{\bq}{\mathbf{q}}

\newcommand*{\bx}{\mathbf{x}}

\newcommand*{\fnl}{f_{\rm NL}}

\newcommand*{\deltam}{\delta_{\rm M}}
\newcommand*{\deltaR}{\delta_{g,R}}

\newcommand*{\bea}{\begin{eqnarray}}
\newcommand*{\eea}{\end{eqnarray}}
\newcommand*{\be}{\begin{equation}}
\newcommand*{\ee}{\end{equation}}

\title {The constraining power of the Marked Power Spectrum: an analytical study}

\author[a,b]{Marco Marinucci}
\author[c]{, Gabriel Jung}
\author[a,b]{, Michele Liguori}
\author[a,b]{, Andrea Ravenni}
\author[a,b]{, Francesco Spezzati}
\author[d,e]{, Adam Andrews}
\author[d,e,f]{, Marco Baldi}
\author[g, h]{William~R.~Coulton}
\author[3,4,5]{, Dionysios Karagiannis}
\author[6]{, Francisco Villaescusa-Navarro}
\author[7,8,9,6]{, Benjamin~D.~Wandelt}

\affiliation[a]{Dipartimento di Fisica e Astronomia “G. Galilei”, Universit`a degli Studi di Padova, via Marzolo 8,
I-35131, Padova, Italy}
\affiliation[b]{INFN, Sezione di Padova, via Marzolo 8, I-35131, Padova, Italy}
\affiliation[c]{Université Paris-Saclay, CNRS, Institut d’Astrophysique Spatiale, 91405 Orsay, France}
\affiliation[d]{INAF/OAS Bologna, via Piero Gobetti 101, I-40129 Bologna, Italy}
\affiliation[e]{INFN, Sezione di Bologna, via Irnerio 46, I-40126 Bologna, Italy}
\affiliation[f]{Dipartimento di Fisica e Astronomia, Alma Mater Studiorum - University of Bologna, Via Piero Gobetti 93/2, 40129 Bologna BO, Italy}
\affiliation[g]{Kavli Institute for Cosmology Cambridge, Madingley Road, Cambridge CB3 0HA, UK}
\affiliation[h]{DAMTP, Centre for Mathematical Sciences, University of Cambridge, Wilberforce Road, Cambridge CB3 OWA, UK}
\affiliation[3]{Dipartimento di Fisica e Scienze della Terra, Universit{\`a} degli Studi di Ferrara, via Giuseppe Saragat 1, 44122 Ferrara, Italy}
\affiliation[4]{INFN, Sezione di Ferrara, via Giuseppe Saragat 1, 44122 Ferrara, Italy}
\affiliation[5]{Department of Physics \& Astronomy, University of the Western Cape, Cape Town 7535, South Africa}
\affiliation[6]{Center for Computational Astrophysics, 160 5th Avenue, New York, NY, 10010, USA}
\affiliation[7]{Department of Physics and Astronomy, Johns Hopkins University, 3400 North Charles Street, Baltimore, MD, 21218, USA}
\affiliation[8]{Department of Applied Mathematics and Statistics, Johns Hopkins University, 3400 North Charles Street, Baltimore, MD, 21218, USA}
\affiliation[9]{CNRS \& Sorbonne Universit\'{e}, Institut d’Astrophysique de Paris (IAP), UMR 7095, 98 bis bd Arago, F-75014 Paris, France}

\emailAdd{marco.marinucci@unipd.it}

\abstract{The marked power spectrum - a two-point correlation function of a transformed density field - has emerged as a promising tool for extracting cosmological information from the large-scale structure of the Universe. In this work, we present the first comprehensive analytical study of the marked power spectrum's sensitivity to primordial non-Gaussianity (PNG) of the non-local type. We extend previous effective field theory frameworks to incorporate PNG, developing a complete theoretical model that we validate against the Quijote simulation suite. Through a systematic Fisher analysis, we compare the constraining power of the marked power spectrum against traditional approaches combining the power spectrum and bispectrum (P+B). We explore different choices of mark parameters to evaluate their impact on parameter constraints, particularly focusing on equilateral and orthogonal PNG as well as neutrino masses. Our analysis shows that while marking up underdense regions yields optimal constraints in the low shot-noise regime, the marked power spectrum's performance for discrete tracers with BOSS-like number densities does not surpass that of P+B analysis at mildly non-linear scales ($k \lesssim 0.25 \,h/\text{Mpc}$). However, the marked approach offers several practical advantages, including simpler estimation procedures and potentially more manageable systematic effects. Our theoretical framework  reveals how the marked power spectrum incorporates higher-order correlation information through terms resembling tree-level bispectra and power spectrum convolutions. This work establishes a robust foundation for applying marked statistics to future large-volume surveys.}

\begin{document}

\maketitle

\section{Introduction}

Galaxy surveys have made remarkable progress in recent years, enabling the study of the large-scale structure (LSS) of the Universe with unprecedented precision. Notably, current and forthcoming LSS observations will give us access to a significant amount of new cosmological information from non-linear scales in the galaxy density field. Extracting this information is however challenging, as it involves either the study of beyond power spectrum statistics or full field-level analysis.
To address this, the most direct approach consists in measuring higher order correlators of the field, starting with the bispectrum (three-point function in Fourier space). This approach has been adopted in many studies, e.g., \cite{2015MNRAS.451..539G, 2021JCAP...07..008G, eftoflss_png_1, eftoflss_png_2, 2023arXiv231015246H, Ivanov:2023qzb, Hahn:2023kky, Byun:2022rvn, Coulton:2022rir,2024JCAP...05..059D, 2023ApJ...948..135J}
Its primary problem is however that the bispectrum remains a nearly lossless statistic only at mildly non-linear scales; extending it to higher orders moments, on the other hand, presents significant numerical and computational challenges, especially when accounting for realistic data analysis and observational issues, such as the necessity to evaluate covariances and include window functions.

For this reason, there has been increasing interest in the study of alternative summary statistics that can efficiently compress information even in the strongly non-linear regime while remaining computationally manageable. A partial list of these summaries includes methods such as the Wavelet Scattering Transform \cite{Blancard:2023iab,2022PhRvD.105j3534V,2024arXiv240718647V,2024PhRvD.109j3503V,2022arXiv220407646E,2024JCAP...07..021P}, k-nearest neighbors \cite{2023MNRAS.519.4856B,2024MNRAS.534.1621C}, skew-spectra \cite{2021JCAP...03..020S,2024PhRvD.109j3528H}, clipping \cite{2013PhRvD..88h3510S} and Minkowski functionals \cite{2021MNRAS.508.3771L,2023JCAP...09..037L}.
So far, these summaries have been either explored through Fisher analysis or used as inputs for likelihood-free inference.

In this work, we focus on another promising statistic that has received significant attention in the literature: the marked power spectrum. This is defined as the two-point correlator of a weighted version of the density field, where the weight (``mark") can represent galaxy properties, halo merger history or - as considered in this work -- the average density of the surrounding environment.

The marked power spectrum was first introduced in \cite{Stoyan}. In the context of Cosmology, it was recently pointed out in \cite{White_mark} that a density-weighted mark could be a powerful probe of Modified Gravity. Its effectiveness at constraining many cosmological parameters -- and notably neutrino masses -- was shown by the authors of \cite{Massara:2020pli, Massara:2022zrf} in a simulation-based study. Its sensitivity to primordial non-Gaussianity was instead recently pointed out in \cite{Jung_Quij}.

Previous works \cite{White_mark, Massara:2022zrf, Philcox_1, Philcox_2, 2024arXiv240917133E} emphasized two main factors that explain why the marked power spectrum is so efficient at extracting cosmological information, compared to the standard power spectrum. The former is that, by correlating the field with the environment, the marked power spectrum is effectively incorporating information from higher-order correlation functions. The latter is that the mark can be chosen to up-weight low-density regions. These carry significant information but, at the same time, give very little contribution to the conventional power spectrum. 

The primary aim of this work is to shed further insight into these issues by significantly extending previous analytical studies of the marked power spectrum in several directions.

First, we extend the EFTofLSS-based formalism introduced by \cite{Philcox_1}, developing for the first time a theoretical framework that incorporates primordial non-Gaussianity (PNG) into the perturbative description of the marked power spectrum.
Additionally, we conduct a Fisher matrix analysis to give an analytical quantitative assessment of the marked power spectrum's sensitivity to various parameters, including PNG and neutrino masses; in this assessment, we consider different choices of the density-weight parameters of the mark in order to verify the impact of under/over-weighting low-density regions. Finally, we make a detailed comparison between the constraining power of the marked power spectrum and that of a combination of the conventional power spectrum and bispectrum ($P+B$). In previous works, the marked 2-point statistic was typically compared to the standard power spectrum, but $P+B$ looks like a more natural benchmark, considering that the marked field inherently includes higher-order information, as noted earlier. This will be made more transparent in our analytical model, which displays new terms in the one-loop power spectrum of the marked fields that are very similar to the tree-level bispectrum and to the convolution of linear power spectra.

In the current study, we focus specifically on the form of the mark introduced in \cite{White_mark}, which is widely used in cosmological studies on this topic. Recently, \cite{2024arXiv240905695C} highlighted that this ansatz can be optimized in a parameter-dependent manner to enhance estimation efficiency. We leave this optimization — especially in the context of PNG analysis, which is a primary focus of our study — for future work. Additionally, we restrict our analysis to biased tracers in real space, with an extension to redshift space also planned for a forthcoming publication.

The paper is structured as follows. In Section 2, we present our theoretical modeling of the marked power spectrum, developing the perturbative framework and including the effects of primordial non-Gaussianities. We discuss the necessary counterterms, shot noise contributions, and IR-resummation techniques required for a complete theoretical description. Section 3 contains our main results, beginning with a detailed comparison of our theoretical predictions against the Quijote simulations. We then present Fisher forecasts comparing the constraining power of the marked power spectrum to standard P+B analyses, examining both the dark matter field and biased tracers. We explore how different choices of mark parameters affect constraints on neutrino masses and non-local PNG, and investigate the role of bias parameters in degrading these constraints. In Section 4, we discuss the implications of our findings, comparing with previous work and highlighting the practical advantages and limitations of the marked power spectrum approach. We also outline several promising directions for future work, including the extension to redshift space and the possibility of optimizing the mark for specific cosmological parameters. We present our conclusions in Section 5.

\section{Theory modeling for the Marked Spectrum}
The standard statistic employed in Large Scale Structure analysis is the power spectrum of the galaxy number density perturbation\footnote{Throughout this work, we will always refer to galaxies, but the present approach refers to any biased tracer of the dark matter density field.}
\begin{equation}
    n_g(\bx) \equiv \bar{n}_g(1 + \delta_g(\bx))\,,
    \label{eq:ng}
\end{equation}
where $\bar{n}_g\equiv \langle n_g(\bx)\rangle$ is the mean galaxy density and $\delta_g(\bx)$ is the usual overdensity field. In this work, we are interested in studying an alternative summary statistic to extract more information from the galaxy clustering, the \textit{Marked Power Spectrum} (see refs.~\cite{Jung_Quij, Beyond-2pt} for other examples of summary statistics), in which the galaxy number density of eq.~\ref{eq:ng} is weighted by a (space-dependent) mark 
\begin{equation}
    \rho_{\rm M} (\bx) = m(\bx) n_g(\bx) = m(\bx)\bar{n}[1 + \delta_g(\bx)]\,.
\end{equation}
The mark function usually adopted in literature is~\cite{White:2016yhs, Aviles_MPS, Massara:2020pli, Philcox_1, Philcox_2}
\begin{equation}
    m(\bx) = \left(\frac{1 +\delta_s}{1 +\delta_s + \delta_{g, R}(\bx)}\right)^p\,,
    \label{eq:mark_here}
\end{equation}
which is a function of the smoothed galaxy overdensity field, $\deltaR(\bx)$\footnote{We will use a Gaussian smoothing function (in Fourier space) which reads
\begin{equation}
    W_R(k) = e^{-\frac{k^2 R^2}{2}}\,,\quad\,\deltaR(\bk) = W_R(k)\delta_g(\bk)\,.
\end{equation}
}, and $R$, $\delta_s$, and $p$ are hyper-parameters of the model that can be used to enhance underdense ($p>0$) and overdense ($p<0$) regions. Marking the overdensity field with this non-linear function of the smoothed field will make the marked density field more non-Gaussian, allowing us to access information on the large-scale structure distribution that is usually included in higher-order correlation functions. 

\subsection{Eulerian perturbation theory}
By defining the mean mark $\bar{m}$ as
\begin{equation}
    \langle \rho_{\rm M}(\bx)\rangle = \langle m(\bx)n_g(\bx) \rangle \equiv \bar{m} \bar{n}
\end{equation}
we can consider the marked overdensity field 
\begin{equation}
    \delta_M(\bx) \equiv \frac{\rho_M(\bx) - \langle\rho_M\rangle}{\langle \rho_M\rangle} = \frac{m(\bx)}{\bar{m}}[1 + \delta_g(\bx)] - 1\,,
    \label{eq:mark_1}
\end{equation}
and expand it perturbatively in the non-linear fields $\delta_g$ and $\delta_{g,R}$. Expanding the mark, we first obtain
\begin{align}
    m(\bx) &= \left(\frac{1 + \delta_s}{1 + \delta_s + \deltaR}\right)^p = \sum_{n = 0}^{\infty} (-1)^n\frac{p(p+1)\dots(p+n-1)}{n!}\frac{\deltaR^n}{(1 + \delta_s)^n} \nonumber\\
    &\equiv \sum_{n=0}^{\infty} (-1)^n C_n\deltaR^n(\bx)\,,
    \label{eq:m_ser}
\end{align}
with
\begin{equation}
    C_n \equiv \frac{p(p+1)\dots (p+n-1)}{n!}\frac{1}{(1 + \delta_s)^n}\,.
\end{equation}
For biased tracers, such as galaxies, we need to be careful about the convergence of the series in \eqref{eq:m_ser}. 
Taking a large enough $R$, we know that the time evolution of the smoothed field is $\deltaR(z) \simeq b_1(z) D(z) \delta_0$, with $b_1$ and $D$ being the linear bias and the growth factor; the convergence of the series is guaranteed simply by $|\deltaR(\bx)/(1 + \delta_s)| < 1$, and, noting that the fluctuation of the smoothed field is $\sigma_{RR}^2(z) \equiv \langle\deltaR^2(\bx)\rangle$, we expect convergence if
\begin{equation}
    \sigma_{RR}(0) < \frac{1 + \delta_s}{b_1(z) D(z)}\,.
\end{equation}

We can proceed by expanding perturbatively as
\begin{equation}
    \delta_g(\bx) = \sum_{n = 0}^{\infty}\delta_g^{(n)}(\bx)\,,\quad\quad \deltaR(\bx) = \sum_{n = 0}^{\infty}\deltaR^{(n)}(\bx)\,,
    \label{eq:exp_1}
\end{equation}
where, in Fourier space, we have $\deltaR^{(n)}(\bk) = W_R(k)\delta_g^{(n)}(\bk)$. Now, inserting~\eqref{eq:exp_1} inside the marked density field definition, eq.~\eqref{eq:mark_1}, we obtain
\begin{align}
    \bar{m}(\deltam (\bx) -1) &= m(\bx) (1 + \delta_g(\bx)) \nonumber\\
    & = \sum_{n=0}^{\infty}(-1)^n C_n \left(\sum_{l=0}^{\infty}\deltaR^{(l)}(\bx)\right)^n\left[1 + \sum_{j=1}^{\infty}\delta_g^{(j)}(\bx)\right] \nonumber\\
    & \equiv 1 + \sum_{n=0}^{\infty} \deltam^{(n)}(\bx) \,,
\end{align}
from which we obtain
\begin{align}
    &\deltam^{(1)}(\bx) = C_0\delta_g^{(1)}(\bx) - C_1 \deltaR^{(1)}(\bx)\,,\\
    &\deltam^{(2)}(\bx) = C_0\delta_g^{(2)}(\bx) - C_1\deltaR^{(1)}\delta_g^{(1)}(\bx) - C_1\deltaR^{(2)}(\bx) + C_2\deltaR^{(1)}\deltaR^{(1)}(\bx)\,,\\
    &\deltam^{(3)}(\bx) = C_0\delta_g^{(3)}(\bx) - C_1\deltaR^{(1)}\delta_g^{(2)}(\bx) - C_1 \deltaR^{(2)}\delta_g^{(1)}(\bx) - C_1\deltaR^{(3)}(\bx)\\
    &\quad\quad\quad\quad\,C_2\deltaR^{(1)}\deltaR^{(1)}\delta_g^{(1)}(\bx) + 2C_2\deltaR^{(1)}\deltaR^{(2)}(\bx) - C_3 \deltaR^{(1)}\deltaR^{(1)}\deltaR^{(1)}(\bx)\,.\nonumber
\end{align}
To proceed, we expand the various fields involved into power of the linear matter perturbation field $\delta^{(1)}(\bk)$. For biased tracers in real space, we have
\begin{equation}
    \delta_g^{(n)}(\bk, z) = \mathcal{I}_{\bk,\bq_1, \dots, \bq_n}K_n(\bq_1, \dots, \bq_n; z)\delta^{(1)}(\bq_1)\dots \delta^{(1)}(\bq_n)\,,
    \label{eq:pt_exp}
\end{equation}
where we have defined
\begin{equation}
    \mathcal{I}_{\bk,\bq_1, \dots, \bq_n}\equiv \int\frac{d^3\bq_1}{(2\pi)^3}\dots \int\frac{d^3\bq_n}{(2\pi)^3}(2\pi)^3\delta_D(\bk - \bq_{1\dots n})\,,
\end{equation}
and $K_n$ are the real space kernels for a generic biased tracer~\cite{Bernardeau_PT, Ivanov_PS, boot}. Analogously, we can define new kernels for the marked field as
\begin{equation}
    \deltam^{(n)}(\bk) = \mathcal{I}_{\bk, \bq_1, \dots, \bq_n}H_n(\bq_1, \dots, \bq_n; z)\delta^{(1)}(\bq_1)\dots \delta^{(1)}(\bq_n)\,,
\end{equation}
with
\begin{align}
    H_1(\bk) &= C_{\deltam}(k) K_1(\bk)\,,\nonumber\\
    H_2(\bk_1, \bk_2) &= C_{\deltam}(k) K_2(\bk_1, \bk_2) + C_{\deltam^2}(k_1, k_2)K_1(\bk_1)K_1(\bk_2)\,, \label{eq:kern}\\
    H_3(\bk_1, \bk_2, \bk_3) &=  C_{\deltam}(k) K_3(\bk_1, \bk_2, \bk_3)  + 2 C_{\deltam^2}(k_1, k_{23})K_1(\bk_1)K_2(\bk_2, \bk_3) \nonumber\\
    &\quad+C_{\deltam^3}(k_1, k_2, k_3) K_1(\bk_1)K_1(\bk_2)K_1(\bk_3)\,, \nonumber
\end{align}
where the $H_3$ still needs to be symmetrized of the three momenta. We have defined the functions
\begin{align}
    C_{\deltam}(k) = &\, C_0 - C_1 W_R(k)\,, \label{eq:Cd1}\\
    C_{\deltam^2}(k) = &\, C_2 W_R(k_1)W_R(k_2) - \frac{C_1}{2}[W_R(k_1) + W_R(k_2)]\,,\label{eq:Cd2}\\
    C_{\deltam^3}(k_1, k_2, k_3) = &\, -C_3 W_R(k_1)W_R(k_2)W_R(k_3) \nonumber\\
    &\, +\frac{C_2}{3}[W_R(k_2)W_R(k_3) + W_R(k_1)W_R(k_3) + W_R(k_1)W_R(k_2)]\,.
    \label{eq:Cd3}
\end{align}
Notice that these functions implicitly contain information about the choice of the adopted mark through the coefficients $C_i$.
Finally, from this expansion, we can compute the power spectrum of the marked density field up to third perturbative order, i.e. the one loop marked power spectrum. We have
\begin{equation}
    \bar{m}^2M(\bk) = \bar{m}^2|\deltam(\bk)|^2\equiv M_{11}(\bk) + 2M_{13}(\bk) + M_{22}(\bk)\,,
\end{equation}
with
\begin{align}
    &M_{11}(\bk) = H_1^2(\bk) P_L(k)\,,\\
    &M_{13}(\bk) = 3 H_1(\bk) P_L(k) \int\frac{d^3\bq}{(2\pi)^3} H_3(\bk, \bq, -\bq)P_L(q)\,,\\
    &M_{22}(\bk) = 2\int\frac{d^3\bq}{(2\pi)^3}|H_2(\bk-\bq, \bq)|^2 P_L(q)P_L(|\bk-\bq|)\,,
    \label{eq:ints}
\end{align}
where $P_L$ is the linear power spectrum. By writing explicitly the expression for the $22$ term in eq.~\ref{eq:ints} we obtain 
\begin{align}
    M_{22}(k) =&\,\, 2 C_{\deltam}^2(k)\int_\bq K_2^2(\bk-\bq, \bq) P_{\rm lin}(|\bk - \bq|)P_{\rm lin}(q)\nonumber\\
    &+2\int_\bq C^2_{\deltam^2}(|\bk - \bq|, q)K_1^2(\bk - \bq)K_1^2(\bq)P_{\rm lin}(|\bk - \bq|)P_{\rm lin}(q) \\
    &+ 4 C_{\deltam}(k) \int_\bq C_{\deltam^2}(|\bk - \bq|, q)K_1(\bk - \bq) K_1(\bq)K_2(\bk - \bq, \bq)P_{\rm lin}(|\bk - \bq|)P_{\rm lin}(q)\,.
    \label{eq:M22}
\end{align}
The first line is a simple rescaling of the usual $P_{22}$ term of the galaxy one-loop power spectrum. This piece should contain, at most, the same amount of information of the $P_{22}$ term, modulo the dependence on the scale of the $C_{\deltam}$ function. The additional terms in the second and third line of eq.~\ref{eq:M22}, which are generated by the presence of the smoothed field in the mark function, are convolutions of bispectrum- and (disconnected) trispectrum-like terms with some scale-dependent kernels given by the marking procedure. These terms arise because the marked field is by definition more non-gaussian, and already at the two-point function level, it contains information from higher order correlation functions of the unmarked field.

The kernels $K_n$ that appear in eq.~\ref{eq:kern} need to be evaluated using a given prescription for the bias expansion. We will follow the usual bias expansion described in~\cite{Assassi_bias} up to third order
\begin{equation}
    \delta_g(\bx) = b_1 \delta(\bx) + \frac{b_2}{2}\delta^2(\bx) + b_{\mathcal{G}_2}\mathcal{G}_2(\bx)+  \frac{b_3}{6}\delta^3 (\bx) + b_{\mathcal{G}_3}\mathcal{G}_3(\bx) + b_{\mathcal{G}_2\delta}\mathcal{G}_2(\bx)\delta(\bx) + b_{\Gamma_3}\Gamma_3(\bx)\,,
    \label{eq:bias}
\end{equation}
where we have defined the tidal operators as
\begin{align}
    \mathcal{G}_2(\bx) &\equiv (\partial_i\partial_j\Phi(\bx))^2 - (\partial^2\Phi(\bx))^2\,\\
    \mathcal{G}_3(\bx) &\equiv - \partial_i \partial_j \Phi \partial_j\partial_k \Phi \partial_k \partial_i \Phi - \frac{1}{2}(\partial^2 \Phi)^2 + \frac{3}{2}(\partial_i \partial_j \Phi)^2\partial^2\Phi\,\\
    \Gamma_3(\Phi_g, \Phi_v)(\bx) &\equiv \mathcal{G}_2(\Phi_g)(\bx) - \mathcal{G}_2(\Phi_v)(\bx) \,,
    \label{eq:oper}
\end{align}
with $\Phi, \Phi_v$ being the gravitational and velocity potential perturbations. The only operators that will enter the one-loop power spectrum are the four renormalized operators $\delta$, $[\delta^2]$, $[\mathcal{G}_3]$ and $[\Gamma_3]$, related to the usual fields via
\begin{equation}
    [\delta^2](\bx) = \delta^2 - \sigma^2, \quad [\mathcal{G}_2](\bx) = \mathcal{G}_2(\bx)\,,
\end{equation}
with $\sigma^2$ being the variance of the unsmoothed field
\begin{equation}
    \sigma^2 = \int_\bq P_L(q)\,.
\end{equation}
As pointed out in~\cite{Philcox_1, Philcox_2}, the marked power spectrum requires more caution compared to the standard case as the inclusion of renormalized operators leads to the addition of a zero-lag contribution, which, in Fourier space, manifests as
\begin{equation}
    \delta_g(\bk)\to \delta_g(\bk) - \frac{b_2}{2}\sigma^2(2\pi)^3\delta_D(\bk)\,,
\end{equation}
or, for the marked field at second and third perturbative orders
\begin{align}
    &\deltam^{(2)}(\bk)\to \delta_M^{(2)} - \frac{b_2}{2}\sigma^2 C_{\deltam}(k) (2\pi)^3\delta_D(\bk)\,,\\
    & \deltam^{(3)}(\bk) - b_2 \sigma^2 C_{\deltam^2}(k,0)\delta^{(1)}(\bk)\,.
    \label{eq:renorm}
\end{align}
The inclusion of the third-order contribution is crucial for the small-scale safety of the one-loop marked power spectrum.

In this work, we will compare the information content of a standard analysis that involves galaxy power spectrum and bispectrum against an analysis performed by adopting only two-point correlation functions, namely the one-loop power spectrum and marked power spectrum. The model for the galaxy bispectrum in real space is given by
\begin{equation}
    B(k_1, k_2, k_3) = 2 K_1(\bk_1)K_1(\bk_2)K_2(\bk_1, \bk_2) P_L(k_1)P_L(k_2) + 2 \,\,\text{perms.}\,,
\end{equation}
where we stopped at the tree level. As for the one-loop contribution, one would need to go to the fourth perturbative order. The tree-level approximation has been shown to work properly up to $k_{max}^{B}\simeq 0.08/0.10\, h/\text{Mpc}$~\cite{Ivanov_bisp1, Ivanov_bisp2, Philcox_bisp_BOSS}, while the one-loop is needed to extend the maximum $k$ and get unbiased results~\cite{Philcox_bisp_oneloop}.

\subsection{Counterterms, shot noise and IR-resummation}
The perturbative results presented in the previous section need to be corrected in order to achieve better agreement with smaller, more non-linear scales. 
The Effective Field Theory of Large Scale Structure (EFTofLSS), originally presented in~\cite{Baumann_EFT, Carrasco_EFT, Pietroni_EFT}, proposes itself as a completion treatment of standard perturbation theory, addressing its main issues: UV divergence and the treatment of imperfections in the cosmological fluid. For the standard one-loop power spectrum in real space, this is achieved with the introduction of a third-order counterterm, which, in Fourier space, reads $\delta^{(ct)}(\bk) = - c_s^2k^2 \delta^{(1)}(\bk)$. This term accounts for the backreaction of small-scale physics on long-wavelength modes. The speed of sound $c_s^2$ is a free parameter of the theory and will be fitted with observational data. For the marked power spectrum, it was shown that this term is enough to capture the UV divergences in real space; all other terms are manifestly convergent for hard loop momenta $p\gg k$ due to the presence of smoothing windows that depend on the physical scale $R$, used in the definition of the mark. Ref.~\cite{Philcox_1, Philcox_2} showed that the marked theory does not require additional counterterms, and the relevant counterterm in $M(\bk)$ is 
\begin{equation}
    M_{ct}(\bk) = - 2c_s^2 k^2 C_{\deltam}^2(k)P_L(k)\,,
\end{equation}
with $c_s^2$ being redshift dependent. In the case of biased tracers, additional UV divergences arise
\begin{align}
    &M_{13}(\bk)\supset 2 C_{\deltam}(k)C_{\deltam^2}(k,0)K_1^{2}(\bk) P_L(k)\int_{\bq} \frac{b_2}{2}P_L(q)\,,\\
    &M_{22}(\bk)\supset \frac{b_2^2}{2}C_{\deltam}^2(k)\int_\bq P_L(q)P_L(|\bk - \bq|) = \frac{b_2^2}{2}C^2_{\deltam}(k)\int_\bq [P_L(q) P_L(|\bk - \bq|) - P_L^2(q)] \\
    &\hspace{6.3cm}+ \frac{b_2^2}{2}C_{\deltam}^2(k)\int_\bq P_L^2(q)\,.
\end{align}
The $M_{13}$ contribution vanishes when including the proper renormalized bias operator as in eq.~\ref{eq:renorm}, while the second term contains a divergent part in the last line that can be completely absorbed when considering the shot noise term. Indeed, the stochastic part of the marked power spectrum
\begin{equation}
    M_{\rm stoch}(k) = C_{\deltam}^2(k) P_{\rm shot}\,,
    \label{eq:SN}
\end{equation}
for constant $P_{\rm shot}$ reabsorbs the UV divergence present in the $M_{22}$ term. 

For the infrared resummation, that accounts for the correct treatment of long-wavelength (IR) displacements on the baryon acoustic peak~\cite{Senatore_IR, Baldauf_IR}, we will follow the approximation of~\cite{Philcox_1, Philcox_2}: for the terms in the marked spectrum involving $C_{\deltam}^2(k)P_{NL}(k)$ we replace $P_{NL}$ with its IR-resummed form, while for the other terms that involve the linear power spectrum $P_L(k)$, we replace it with the IR-resummed one. The IR-resummed power spectrum for matter has the form~\cite{Ivanov_IR}
\begin{align}
    P_{\rm LO}^{\rm IR-res}(k) &= P_{\rm L}^{nw}(k) + e^{- k^2 \Sigma^2}P_{\rm L}^{w} \\
    P_{\rm NLO}^{\rm IR-res}(k) &= P_{\rm L}^{nw}(k) + P_{\rm 1-loop}^{nw}(k) + e^{- k^2 \Sigma^2}\left[(1 + k^2 \Sigma^2) P_{\rm L}^{w} +  P_{\rm 1-loop}^{w}\right]\,,
\end{align}
where $nw$ and $w$ parts are the broadband (no-wiggle) and the oscillatory parts of the power spectrum, and $\Sigma$ is the velocity dispersion. Similar relations hold for biased tracers. The effectiveness of this approximation has been shown to fit the simulated data without any residual wiggles appropriately.

Summarizing, the one-loop EFT model for the marked power spectrum of biased tracers in real space has the following form
\begin{equation}
    \bar{m}^2 M(k) = M_{11}(k)+M_{22}(k)+2M_{13}(k)+M_{ct}(k)+M_{\rm stoch}(k)\,,
\end{equation}
which contains the following five parameters\footnote{Notice that we are fixing $b_{\Gamma_3} = 0$, as the power spectrum alone is poorly sensitive to this parameter~\cite{Anton_Euclid, Ivanov_PS}.}
\begin{equation}
    \{b_1, b_2, b_{\mathcal{G}_2}, c_s^2, P_{\rm shot}\}\,.
    \label{eq:list}
\end{equation}
At the perturbative order that we are considering, the galaxy bispectrum presents only the following stochastic terms
\begin{equation}
    B_{\rm stoch}(k_1, k_2, k_3) = P_{\rm shot}\left[P_L(k_1)+P_L(k_2)+P_L(k_3)\right] + B_{\rm shot}\,,
\end{equation}
adding the parameter $B_{\rm shot}$ to the list of varied parameters in eq.~\ref{eq:list}.

\subsection{Primordial non-gaussianities of the non-local type}
The main goal of this work is to investigate the constraining power of marked statistics on primordial non-gaussianities (PNG). In particular, we are interested in primordial non-gaussianities of the non-local type, meaning the non-gaussianities that are suppressed in the squeezed limit and that can arise from quadratic or higher order interaction in the inflationary field~\cite{Cheung_PNG}. 

In this work, we focus on non-local PNG because the potential gain from analyzing the bispectrum is maximized in this scenario, particularly due to the absence of the scale-dependent bias peaked at large scales that appear in local PNGs~\cite{Matarrese:2008nc, Desjacques:2008vf}. Some works have recently explored non-local PNG in the context of machine-learning base approaches to enhance signal extraction~\cite{Giri:2022nzt, Kvasiuk:2024gbz}. Additionally, within the framework of the Cosmological Collider~\cite{Arkani-Hamed:2015bza}, recently adopted on the BOSS survey~\cite{Cabass:2024wob}, several theoretical models involving massive particles during inflation can be effectively approximated by equilateral and orthogonal shapes in the non-local non-Gaussianity regime. This approximation allows for a more accurate representation of interactions in the early universe and provides a pathway for distinguishing between different inflationary models.

It is useful to represent non-local PNG using a linear combination of two shapes, equilateral and orthogonal~\cite{Senatore_PNG}, depending on the triangular configuration over which the bispectrum is peaked. Concretely, the two parameters for the amplitudes of different shapes, $f_{\rm NL}^{\rm eq}$ and $f_{\rm NL}^{\rm or}$, can be directly linked to physical observables of the inflaton field. The Lagrangian of general single-field models in the context of the effective field theory of inflation is given by~\cite{Cheung_PNG, Senatore_PNG}
\begin{align}
    S = \int d^4 x\sqrt{-g}\Bigg[ &- \frac{M_{\rm Pl}^2 \dot{H}}{c_{\pi, s}^2}\left(\dot{\pi}^2 - c_{\pi, s}^2 \frac{(\partial_i\pi)^2}{a^2}\right) - M_{\rm Pl}^2 \dot{H}(1 - c_{\pi, s}^{-2})\dot{\pi}\frac{(\partial_i\pi)^2}{a^2} \label{eq:LagInf}\\
    &+ \left(M_{\rm Pl}^2 \dot{H}(1 - c_{\pi, s}^{-2}) - \frac{4}{3}M_3^{4}\right)\dot{\pi}^3\Bigg]\,, \nonumber
\end{align}
with the scalar perturbation $\pi$ being related to the curvature perturbation $\zeta = - H\pi$. The two leading interactions in eq.~\ref{eq:LagInf} are given by the third order terms $\dot{\pi}(\partial_i\pi)^2$ and $\dot{\pi}^3$, which produce specific bispectra with amplitudes~\cite{Planck_PNG}
\begin{equation}
f_{\rm NL}^{(\dot{\pi}(\partial_i \pi)^2)} = - (85/324)(c_{\pi, s}^{-2} - 1)\,,\quad \text{and}\quad f_{\rm NL}^{\dot{\pi}^3} = - (10/243)(c_{\pi, s}^{-2} - 1)\big[\tilde{c}_3 + (3/2)c_{\pi, s}^2\big]\,,
\end{equation}
where we have defined the dimensionless parameter $\tilde{c}_3$ through $\tilde{c}_3 (c_{\pi, s}^{-2} - 1) = 2 M_3^4 c_{\pi, s}^2 /(\dot{H} M_{\rm Pl}^2)$. The two EFT shapes can be projected onto the equilateral and orthogonal shapes as
\begin{align}
    &f_{\rm NL}^{\rm equil} = \frac{1 - c_{\pi, s}^2}{c_{\pi, s}^2}\big[-0.275 - 0.0780 c_s^2 - 0.53 \tilde{c}_3\big]\,,\\
    &f_{\rm NL}^{\rm ortho} = \frac{1 - c_{\pi, s}^2}{c_s^2}\big[0.0159 - 0.0167 c_{\pi, s}^2 - 0.01113 \tilde{c}_3\big]\,.
\end{align}

The higher order interactions of the inflaton field give rise to a primordial bispectrum for the curvature perturbation $B_\zeta(k_1, k_2, k_3)$ that can be expressed using a linear combination of the orthogonal and equilateral templates. We define
\begin{equation}
    B_\zeta(k_1,k_2,k_3) = \frac{18}{5}\fnl \Delta_\zeta^4\frac{\mathcal{S}(k_1, k_2, k_3)}{k_1^2 k_2^2 k_3^2}\,,
\end{equation}
with the equilateral and orthogonal templates defined as~\cite{Senatore_PNG, Babich_PNG}
\begin{equation}
    \mathcal{S}_{\rm eq}(k_1,k_2,k_3) = \left(\frac{k_1}{k_2} + \text{5 perms.}\right) - \left(\frac{k_1^2}{k_2 k_3}+ \text{2    perms.}\right) - 2\,,
\end{equation}
\begin{equation}
    \mathcal{S}_{\rm ort}(k_1,k_2,k_3) =  (1 + p)\frac{\Delta}{e_3} - p \frac{\Gamma^3}{e_3^2}\,,
\end{equation}
where $p = 8.52587$, $\Delta = (k_T - 2k_1)(k_T - 2 k_2)(k_T - 2 k_3)$
\begin{align}
    k_T = k_1 + k_2 + k_3,  \quad e_2 = k_1 k_2 + k_1 k_3 + k_2 k_3, \quad e_3 = k_1 k_2 k_3, \quad \Gamma = \frac{2}{3} e_2 - \frac{1}{3}(k_1^2 + k_2^2 + k_3^2)\,.
\end{align}
The PNG affects the late-time dark matter distribution, inducing an additional bispectrum term given by
\begin{equation}
    \langle\delta^{(1)}\delta^{(1)}\delta^{(1)}\rangle = \fnl B_{111}(k_1, k_2, k_3)(2\pi)^3\delta_D(\bk_{123})\,
\end{equation}
with
\begin{equation}
    \fnl B_{111}(k_1, k_2, k_3) = \mathcal{T}(k_1)\mathcal{T}(k_2)\mathcal{T}(k_3) B_\zeta(k_1,k_2, k_3)\,,
\end{equation}
where we have introduced the transfer function $\mathcal{T}(k)\equiv \delta^{(1)}(\bk)/\zeta(\bk) = (P_{11}(k)/P_{\zeta}(k))^{1/2}$. Notice that this can be easily calculated via the definition of the power spectrum of initial fluctuations
\begin{equation}
     P_\zeta(k) = \Delta_\zeta^2k^{-3}\left(\frac{k}{k_\ast}\right)^{n_s - 1}\,.
\end{equation}
Analysis of \textit{Planck} data finds $\Delta_\zeta^2\simeq 4.1 \times 10^{-8}$, $n_s \simeq 0.96$ for the pivot scale $k_\ast = 0.05$ Mpc$^{-1}$~\cite{Planck_inflation}. The initial bispectrum also enters the 1-loop galaxy power spectrum through the 1-2 term
\begin{equation}
    \fnl P_{12}(\bk) = 2 \fnl K_1(\bk) \int_\bq K_2(\bk - \bq, \bq)B_{111}(k,q,|\bk - \bq|)\,.
    \label{eq:P12}
\end{equation}
In addition, non-local PNG modulates galaxy formation, which is captured by the scale-dependent galaxy bias
\begin{equation}
    \delta_g = b_1 \delta + \fnl b_\zeta \left(\frac{k}{k_{\rm NL}}\right)^2 \zeta + \text{nonlinear}\,,
\end{equation}
so that the final model for the galaxy power spectra and bispectra in real space is
\begin{equation}
    P_g(\bk) = P_{g, {\rm G}}(k) + \fnl \left( P_{12}(\bk) + 2 b_\zeta K_1(k)\frac{k^2}{k^2_{\rm NL}} \frac{P_L(k)}{\mathcal{T}(k)}\right)\,,
\end{equation}
\begin{equation}
    B_g(\bk_1, \bk_2, \bk_3) = B_{g, {\rm G}}(\bk_1, \bk_2, \bk_3) + \fnl K_1(k_1)K_1(k_2)K_1(k_3)B_{111}(k_1, k_2, k_3)\,,
\end{equation}
where $P_{g,{\rm G}}$ and $B_{g,{\rm G}}$ are the standard Gaussian power spectrum and bispectrum models.

For the marked field, we have the additional term coming from the initial bispectrum contribution
\begin{equation}
    \fnl M_{12}(\bk) = 2 \fnl H_1(\bk) \int_\bq H_2(\bk - \bq, \bq)B_{111}(k, q, |\bk - \bq|)\,,
\end{equation}
so that the marked power spectrum and bispectrum for galaxies in real space are
\begin{equation}
    M_g(\bk) = M_{g, {\rm G}}(\bk) + \fnl \left(M_{12}(\bk) + 2 b_\zeta K_1(k)\frac{k^2}{k^2_{\rm NL}} C_{\deltam}^2(k)\frac{P_L(k)}{\mathcal{T}(k)}\right)
\end{equation}
\begin{equation}
    BM_g(\bk_1,\bk_2, \bk_3) = BM_{g, {\rm G}}(\bk_1,\bk_2, \bk_3)  + H_1(\bk_1)H_1(\bk_2)H_1(\bk_3)\fnl B_{111}(k_1, k_2, k_3)\,,
\end{equation}
with 
\begin{equation}
    BM_{g, {\rm G}} (\bk_1,\bk_2, \bk_3) = 2 H_1(\bk_1)H_1(\bk_2)H_2(\bk_1, \bk_2) P_L(k_1)P_L(k_2) + \text{2 perms.}\,.
\end{equation}
Writing down $M_{12}(\bk)$ explicitly:
\begin{equation}
\begin{gathered}
M_{12}(\bk)=2C_{\delta_M}(k)^2b_1\int_\bq\bigg[b_1F_2(\bk-\bq,\bq)+\frac{b_2}{2}+b_{\mathcal{G}_2}\sigma^2(\bk-bq,\bq)\bigg]B_{111}(k,|\bk - \bq|,q))\\
+2C_{\delta_{M}}(k)b_1^3\int_\bq C_{\delta_{M}}^2(|\bk-\bq|,q)B_{111}(k,|\bk - \bq|,q)\,.
\end{gathered}
\end{equation}
Through this work, we will assume that the following relation holds for the primordial non-Gaussian bias and the linear bias
\begin{equation}
    b_\zeta = \frac{18}{5} \delta_c(b_1 - 1)\,,
    \label{eq:bzeta}
\end{equation}
motivated by the peak-background split argument~\cite{Schmidt_NLPNG}, with $\delta_c$ being the critical overdensity for the spherical collapse.

It is worth noticing that, at the power spectrum level, the PNG contributions are entered only at the one-loop level. Moreover, as shown in~\cite{eftoflss_png_1}, the effect of equilateral and orthogonal non-Gaussianities is very degenerate with the second order biases $b_2$ and $b_{\mathcal{G}_2}$. These two effects conspire to make power spectrum only analyses very inefficient in measuring non-local PNG, making the inclusion of the bispectrum essential in order to to constrain them. The marked power spectrum seems very promising in this direction by bringing information from higher-order functions back to a two-point function, as also shown in~\cite{2024arXiv240917133E}. This could naively explain the results of ref.~\cite{Jung_Quij}, where the authors find that the marked power statistic helps in constraining non-local PNG much more than that of the local type.

\subsection{Low-$k$ behavior}
Ref.~\cite{Philcox_2} showed that, differently from the standard perturbative treatment of the power spectrum, when considering marked statistics, the perturbative series converges more slowly, and loop terms can give non-zero contributions also to linear scales. The authors came up with a low-$k$ correction for the marked power spectrum that, in real space, is
\begin{equation}
    M|_{\text{low}-k}(k) = C_{\deltam}^2(k) \left[a_0 P_L(k) + c_0\right]\,,
\end{equation}
with the addition of two more parameters. We can already exclude the second one since it is completely degenerate with the shot noise term, leaving only the free parameter $a_0$. One should also vary this parameter, but we will fix it for the following reasons. In~\cite{Philcox_2}, they showed that for biased tracers in real space this should not matter much, which also confirmed by our fit against simulations shown in section~\ref{sec:sims}; in this work we want to explore what is the actual gain of the marked power spectrum compared to standard perturbative analyses and fixing $a_0 = 0$ represents the optimistic scenario where we can completely trust our theory model and exploit all of its information content.

\section{Results}
\subsection{Comparison to simulations}
\label{sec:sims}
\begin{figure}
    \centering
    \includegraphics[width=0.325\linewidth]{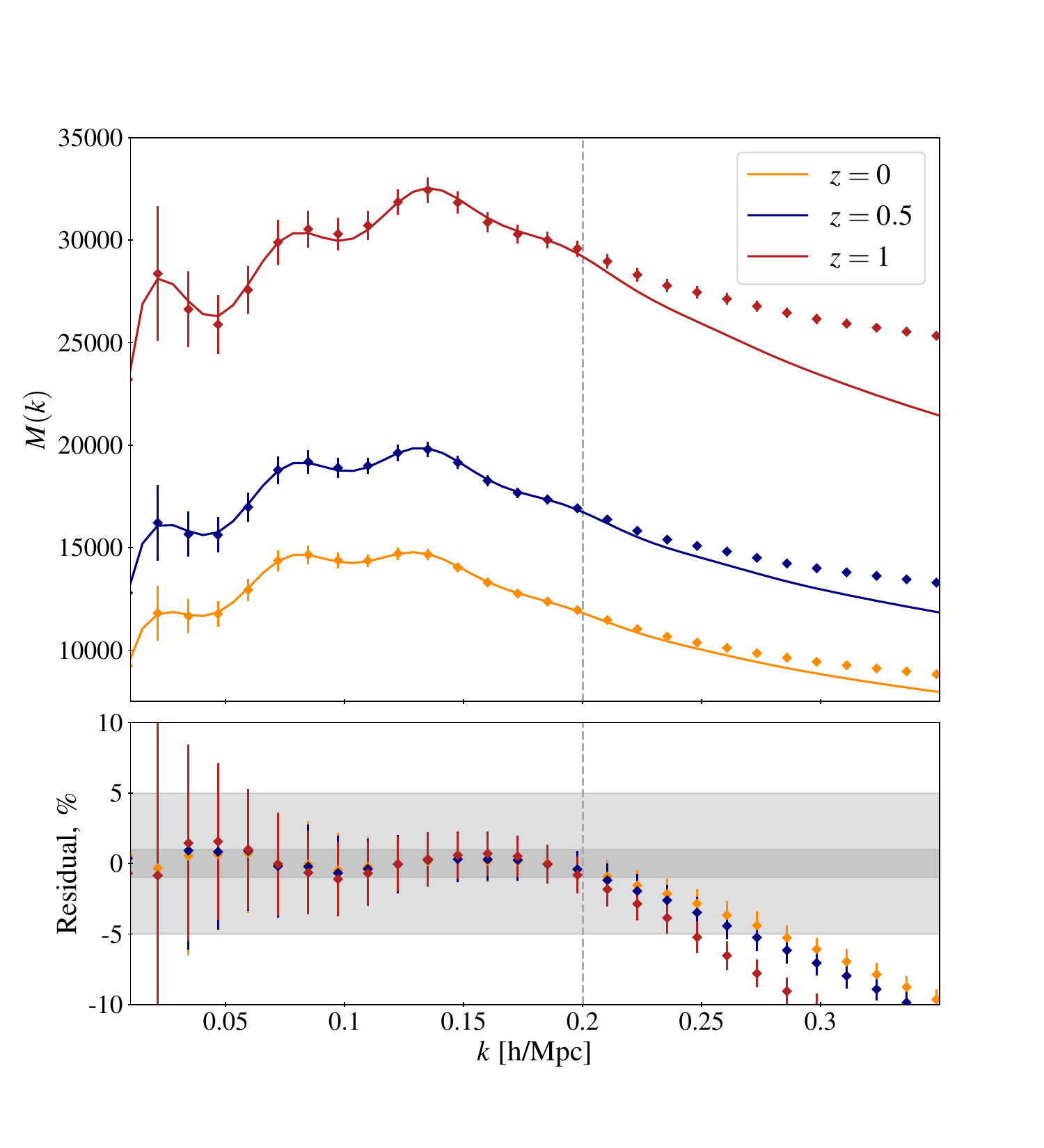}
    \includegraphics[width = 0.325\linewidth]{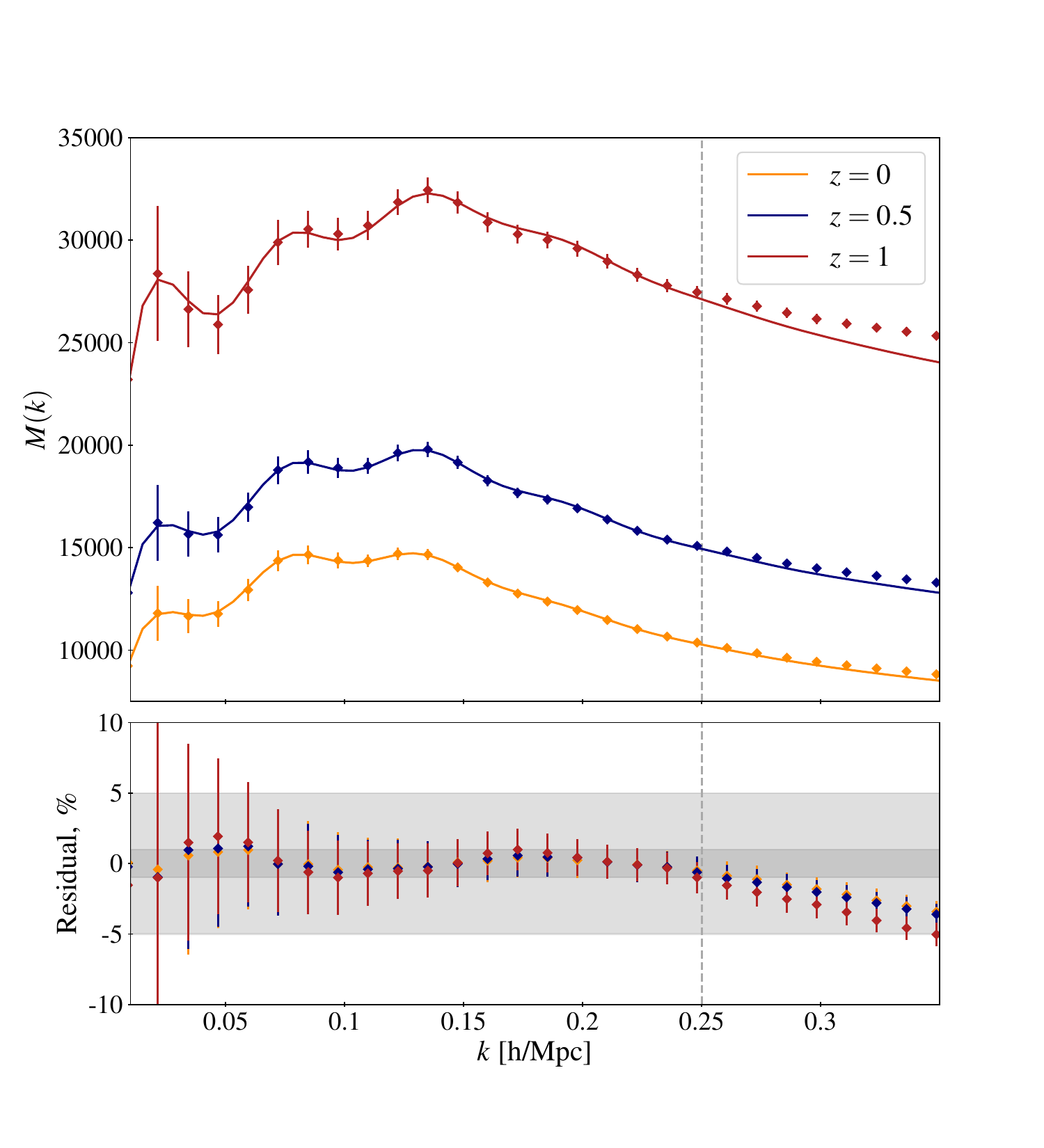}
    \includegraphics[width = 0.325\linewidth]{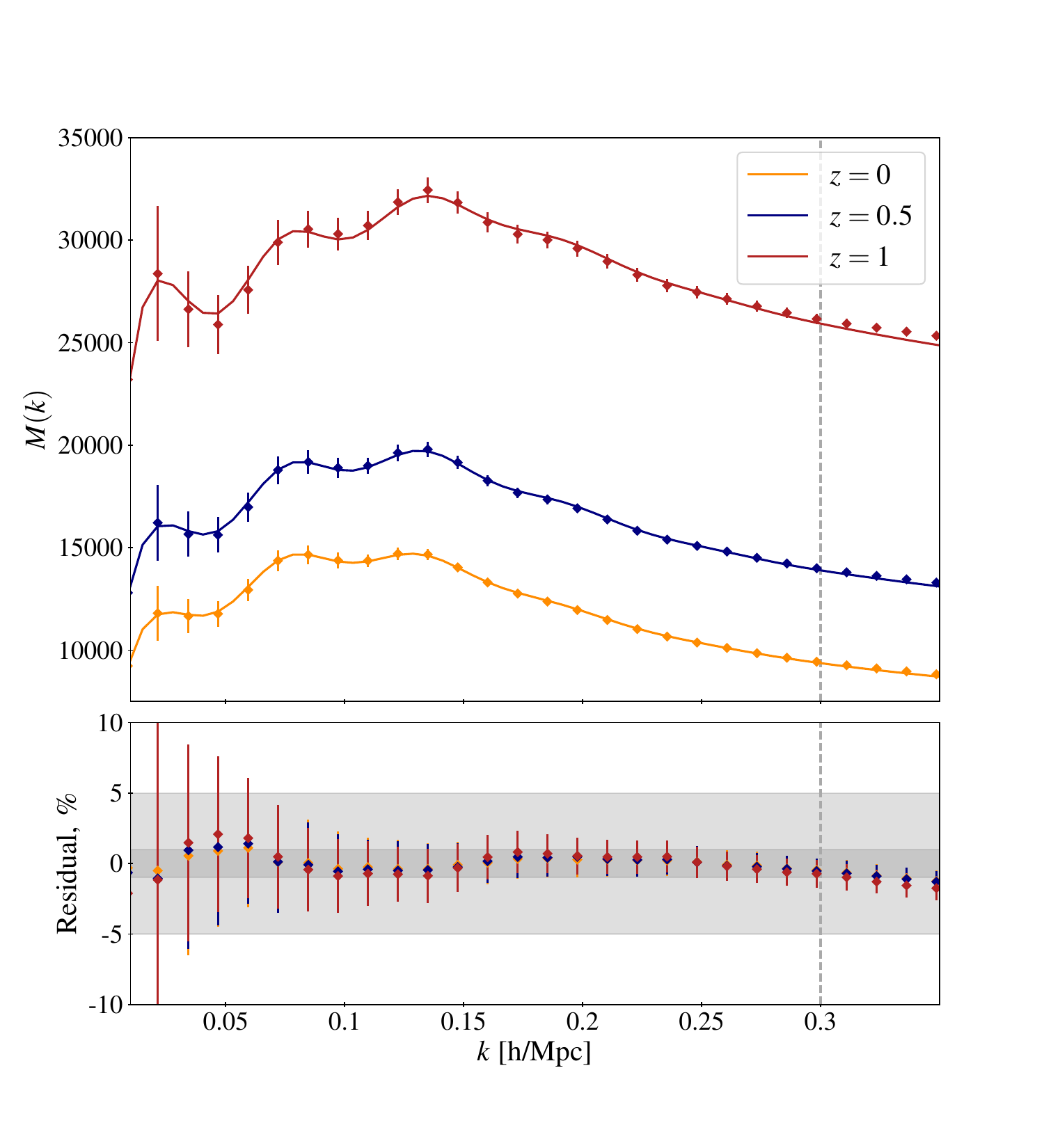}
    \caption{Results of the fitting of the marked power spectrum from the Quijote-PNG simulations with equilateral PNG, $f_{\rm NL}^{\rm eq}  = 100$ for different redshifts. The left plot is obtained fitting the simulations up to $k_{\rm max}^{M} = 0.2 h/\textbf{Mpc}$, the central one with $k_{\rm max}^{M} = 0.25 h/\textbf{Mpc}$ and the right one with $k_{\rm max}^{M} = 0.3 h/\textbf{Mpc}$. The diamonds represent the mean marked power spectrum from the simulations, while the solid lines are the results obtained from the EFT fit.}
    \label{fig:Quij_fit}
\end{figure}

Before proceeding with the Fisher forecast, we compare our predictions with the Quijote~\cite{Villaescusa-Navarro:2019bje} and  Quijote-PNG~\cite{Coulton:2022qbc} simulations. These are N-body simulations of volume 1 (Gpc/\textit{h})$^3$, containing $512^3$ particles each, and run using the TreePM code GADGET-III from initial conditions generated at $z=127$ be the codes 2LPTIC~\cite{Crocce:2006ve} and 2LPTPNG~\cite{Scoccimarro:2011pz, Coulton:2022qbc}, for the simulations with and without PNG respectively. We will compare our PT predictions with real space dark matter halos identified in each simulation by the standard Friends-of-friends algorithm~\cite{Davis:1985rj} by setting the linking length parameter to $b = 0.2$ and considering halos with more the 20 dark matter particles. The fiducial cosmology is $\{\sigma_8, n_s, \Omega_b, \Omega_m, h\}  =  \{0.834, 0.9624, 0.049, 0.3175, 0.6711\}$.
We compare with the simulations at redshifts $z = \{0, 0.5, 1\}$ and mark parameters $\{p = 1, R=15 \,h/\text{Mpc}, \delta_s = 0.25\}$ to test our PT model and numerical implementation by fixing the cosmological parameters to their fiducial values and fitting the bias, small-scales and stochastic parameters $\{b_1, b_2, b_{\mathcal{G}_2}, c_s^2, P_{\rm shot}\}$ to the simulation power spectrum. The results of this fit are shown in figure~\ref{fig:Quij_fit}, where we display the mean marked power spectrum obtained from the PNG simulations fitted up to $k_{\rm max} = \{0.20, 0.25, 0.30\} \,h/\text{Mpc}$. To perform this comparison, we estimate the data covariance numerically using the scattering among the different realizations of the simulations with the same cosmology.
We observe that the theoretical prediction with the EFT of LSS is in excellent agreement with the simulations, within $1\,\%$, up to the maximum $k_{\rm max}$ adopted for the fitting and for different redshifts. These results confirm again the additional terms introduced in~\cite{Philcox_2} to control the low-$k$ behavior of the marked power spectrum are not needed for biased tracers in real space, also in the presence of non-local PNG. Moreover, we do not see any residual wiggles, corroborating the validity of the approximation adopted for the IR-resummation.

\subsection{Analysis specifics}
We are now ready to study the amount of information accounted for by the marked power spectrum and compare it with a standard power spectrum plus bispectrum analysis. This can be readily addressed  by assuming, at leading order, that the statistics considered follow a multivariate Gaussian distribution and compute the Fisher matrix defined as
\begin{equation}
    F_{\mu\nu} = \sum_{i,j}\frac{\partial D_i}{\partial \theta_\mu}C^{-1}_{ij}\frac{\partial D_j}{\partial \theta_\nu}\,,
    \label{eq:Fish}
\end{equation}
where $\mathbf{D} = \{P(\bk), M(\bk), B(\bk_1, \bk_2, \bk_3), \dots\}$ is the data vector, a collection of the different observables employed in the analysis measured at different wavevectors $\bk$. $\theta_\mu$ are the cosmological and EFT parameters that are varied in the analysis, and $C$ is the covariance matrix among the different observables. The Fisher matrix in eq.~\ref{eq:Fish} gives the variance of an optimal unbiased estimator for the parameter $\theta_\mu$, with the marginalized error satisfying $\sigma^2(\theta_\mu)\geq (F^{-1})_{\mu\mu}$: the lower the marginalized error, the better a certain parameter can be constrained with the statistics employed. The statistics that we will consider are the power spectrum, the marked power spectrum and the bispectrum for galaxies in real space. We will compute these statistics by varying the value of $k_{\rm max}$ for P and for M, while we will keep fixed $k_{\rm max}^{B} = 0.10 h$/Mpc for the bispectrum. This has been shown to be a mildly conservative choice as the tree-level model for the bispectrum is valid up to these scales for the volume that we will consider in this work~\cite{Ivanov_PS, Ivanov_bisp1}.

Previous analyses showed that the marked power spectrum always outperforms the standard power spectrum because it is able to keep track of the cosmic web through the non-linear mark in eq.~\ref{eq:mark_here}: different values of the exponent $p$ can enhance voids ($p>0$) or nodes ($p<0$). \cite{Massara:2020pli} finds that positive values of $p$ are optimal to constrain the total neutrino mass and modified gravity theories, given that voids are more sensitive to these two ingredients. In this work, we will perform a Fisher forecast up to mildly non-linear scales using $p=2$ to enhance underdense regions and $p=-1$ to mark the overdense ones to constrain massive neutrinos and non-local-primordial non-gaussianities and compare with the standard analysis of power spectrum combined with bispectrum. The fiducial cosmology adopted here is one of the Quijote simulations for the standard cosmological parameters, with the additional parameters $\{M_\nu, f_{\rm NL}^{\rm eq}, f_{\rm NL}^{\rm or}\} = \{0.06, 0., 0.\}$. We will run the analysis separately for massive neutrinos and primordial non-gaussianities.

For the fiducial of the galaxy biases, we adopt a semi-analytic model of galaxy formation to fix them; specifically, for the linear bias we adopt the following model\footnote{Notice that this model should describe Euclid-like H$\alpha$ targets~\cite{Anton_Euclid}, but we will employ this model for BOSS-like volumes as well since we are only interested in the relative change between different statistics.}
\begin{equation}
    b_1(z) = 0.9 + 0.4 z\,.
\end{equation}
\noindent For the higher order biases we adopt the following fitting formulae~\cite{Bias_Review,Yankelevich_bisp}, obtained from a combination of N-body simulations and halo occupation modeling
\begin{equation}
    b_2(z) = -0.704 - 0.208 z + 0.183 z^2 - 0.00771 z^3\,.
\end{equation}
For $b_{\mathcal{G}_2}$ and $b_{\Gamma_3}$ we use instead the co-evolution model~\cite{Bias_Review, Abidi:2018eyd}, which gives
\begin{equation}
    b_{\mathcal{G}_2} (z) = -\frac{2}{7}(b_1(z) - 1), \quad b_{\Gamma_3}(z) = \frac{23}{42}(b_1(z) - 1)\,.
\end{equation}
For the higher derivative term, we adopt the prescription of~\cite{Anton_Euclid}
\begin{equation}
    c_s = - 25 D^2(z)[\text{Mpc}/h]^2\,,
\end{equation}
with $D(z) = d\ln{\delta_L}/d\ln{a}$ being the growth factor for linear perturbations. Finally, the fiducial values for the stochastic terms are set to the Poisson sampling prediction
\begin{equation}
    P_{\rm shot} = \bar{n}_g^{-1}\, \quad B_{\rm shot} = \bar{n}_g^{-2}\,.
\end{equation}

To compute the one-loop prediction of the standard and the marked power spectrum we implement the public FFTLog-based code CLASS-PT~\cite{Chudaykin:2020aoj}, an extension of the Boltzmann solver CLASS~\cite{Blas:2011rf} that includes the EFTofLSS as non-linear model. For massive neutrino cosmologies we just modify the linear power spectrum and input it inside the one-loop integrals, neglecting the non-linear effects in the perturbative kernels and the scale dependent growth function as this approximation has been shown to perform well for the current constraints on neutrino masses~\cite{Anton_Euclid, Ivanov:2019hqk}.

We perform a forecast for BOSS CMASS2 redshift $z = 0.61$ and total volume $V_{BOSS} = 3.83 $ (Gpc/\textit{h})$^3$, with number density $\bar{n}_g^{BOSS} = 3\times 10^{-4}$ (\textit{h}/Mpc)$^3$.
We consider diagonal Gaussian covariances, after having tested that this approximation correctly reproduces the ones obtained with the Quijote simulations. They are defined as
\begin{equation}
    C_P(k) =\frac{2(2\pi)^3}{V V_s}P^2(k), \quad C_M(k) = \frac{2(2\pi)^3}{V V_s}M^2(k),
\end{equation}
\begin{equation}
C_B(k_1, k2_, k_3) = \frac{(2\pi)^6 s_{123}}{V V_{123}} P(k_1)P(k_2)P(k_3)
\end{equation}
where $V_s = 4 k^2 \delta k$ and $V_{123} = 8\pi^2k_1 k_2 k_3\delta k^3$ are the shell volumes for the power spectrum and the bispectrum, $\delta k =0.01 $ \textit{h}/Mpc, $s_{123}$ is the triangular symmetry factor and $V$ is the volume of the survey. To avoid double counting of the information content in the different spectra, we also include the cross covariance between the marked and the standard power spectrum
\begin{equation}
    C_{PM}(k) = \frac{(2\pi)^3}{V V_s} P(k)M(k)\,.
\end{equation}
In ref.~\cite{Massara:2020pli} it has been shown that the Gaussian approximation for the covariance matrix is particularly good for the marked power spectrum, while for the power spectrum and the bispectrum one should, in principle, also include non-Gaussian terms. These are expected to give a negligible contribution up to the scales considered, although for the bispectrum one should include them to properly estimate the forecasted errorbars, especially for primordial non-Gaussainities~\cite{Floss:2022wkq}.

In standard LSS analyses, the abundance of baryons is usually constrained using BBN priors~\cite{Ivanov_PS}, and hence we decide to fix it to its fiducial value. We vary the cosmological parameters $\{\ln(10^{10}A_s), n_s, \omega_{\rm cdm}, h, M_\nu, f_{\rm NL}^{Equil}, f_{\rm NL}^{Ortho}\}$, with the total neutrino mass and the two PNG parameters varied separately. For galaxies, we fix the third order bias parameter $b_{\Gamma_3}$ to its theoretical prediction, as it is poorly constrained~\cite{Anton_Euclid}, and we vary $\{b_1, b_2, b_{\mathcal{G}_2}, c_s, P_{shot}, B_{shot}\}$.

The list of fiducial parameters is listed in tab.~\ref{tab:fids}

\begin{table}
\centering
\begin{tabular}{||c | c | c | c | c | c | c | c ||} 
 \hline
 $z$ & $b_1$ & $b_2$ & $b_{\mathcal{G}_2}$ & $b_{\Gamma_3}$ & $c_s^{2} [\text{Mpc}/h]^2$ & $P_{\rm shot} [\text{Mpc}/h]^3$  & $B_{\rm shot} [\text{Mpc}/h]^6 \times 10^{-3}$\\
 \hline\hline
 0.61 & 1.14 & -0.38 & -0.041 & 0.079 & 13.21 & 3333 & 11111 \\ 
 \hline
 \end{tabular}
\caption{Table of fiducial values for the redshift, bias parameters, counterterm and shot-noise parameters adopted for the forecast.}
\label{tab:fids}
\end{table}

In the following we will present our results and discuss various effects that contribute to the cosmological parameter measurement from the standard and marked power spectrum and their combination with the bispectrum. We first discuss the results obtained analyzing the dark matter only case and then we will go through the results for galaxies.

\subsection{Dark Matter} 

\begin{figure}[h]
    \centering
    \includegraphics[width=0.325\linewidth]{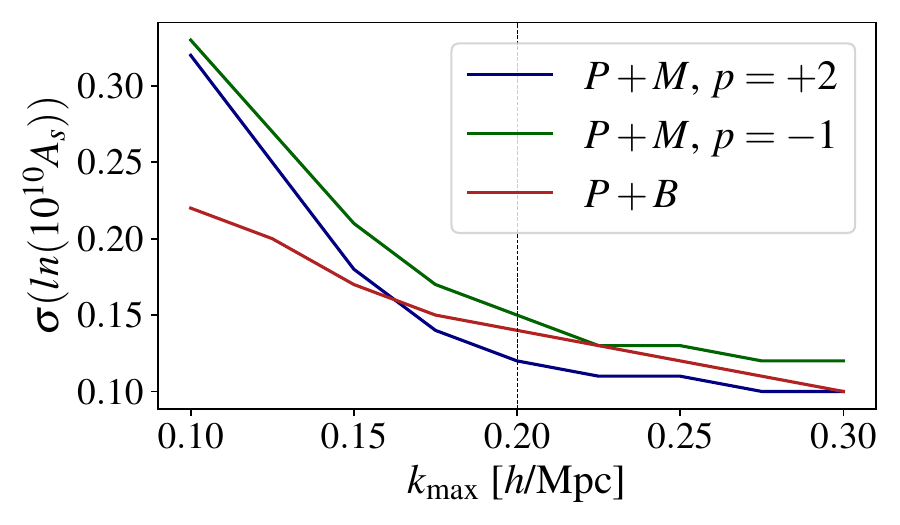}
    \includegraphics[width=0.325\linewidth]{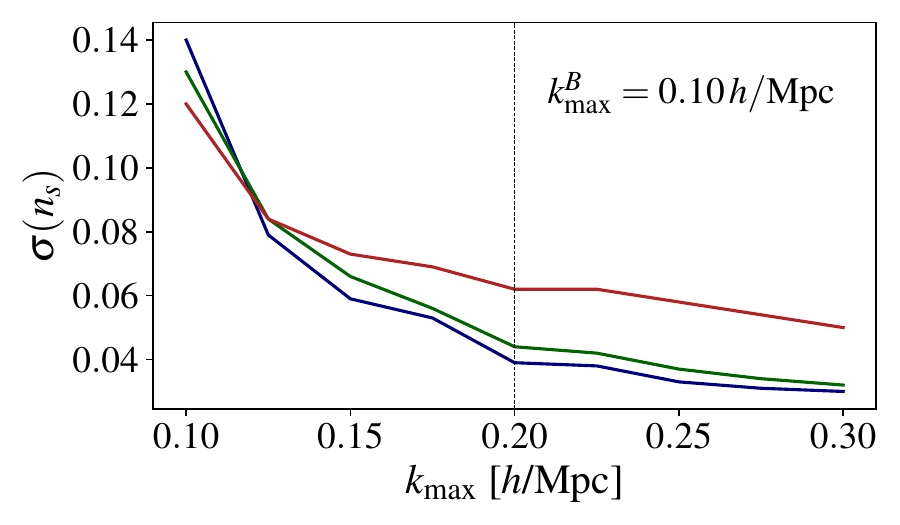}
    \includegraphics[width=0.325\linewidth]{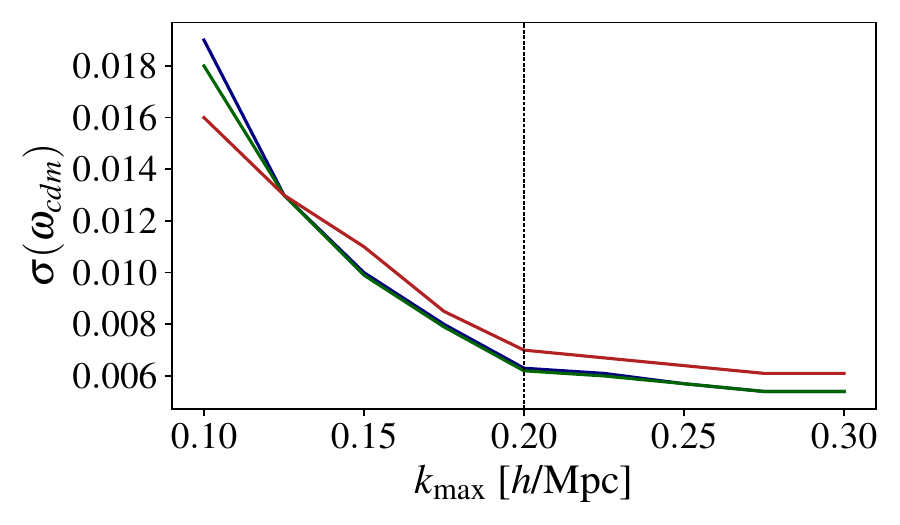}
    \includegraphics[width=0.325\linewidth]{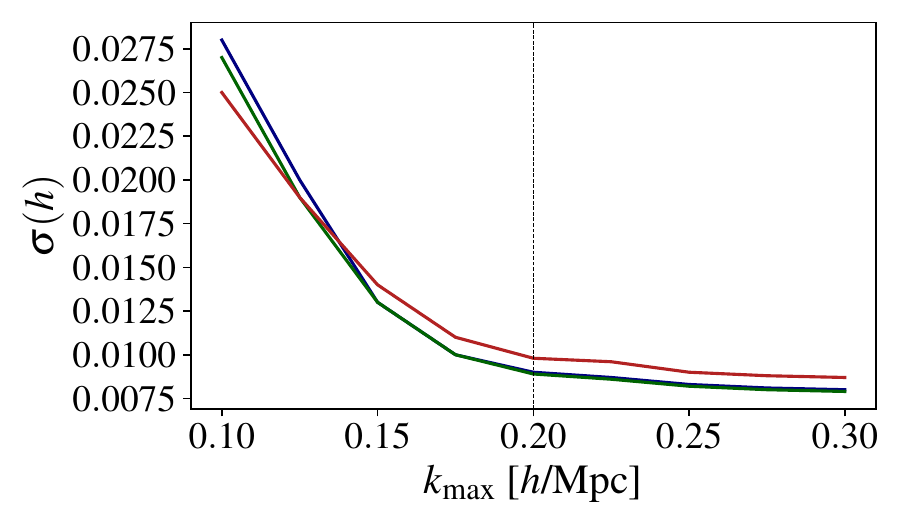}
    \includegraphics[width=0.325\linewidth]{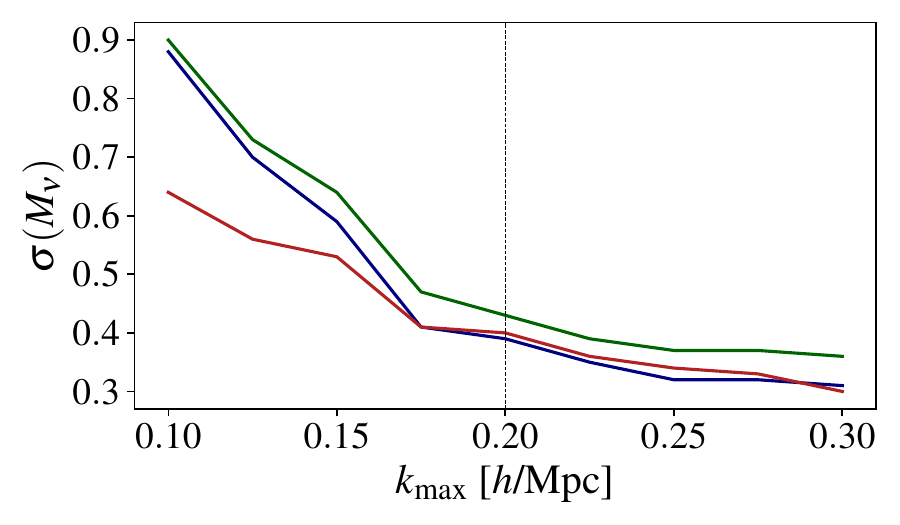}
    \caption{2$\sigma$ forecasted errorbars for dark matter only in presence of massive neutrinos with BOSS CMASS volume and redshift (see main text). The dashed vertical line represents the maximum wave vector adopted in standard analyses~\cite{Ivanov_PS}.}
    \label{fig:matter_Mnu}
\end{figure}

\begin{figure}[h]
    \centering
    \includegraphics[width=0.325\linewidth]{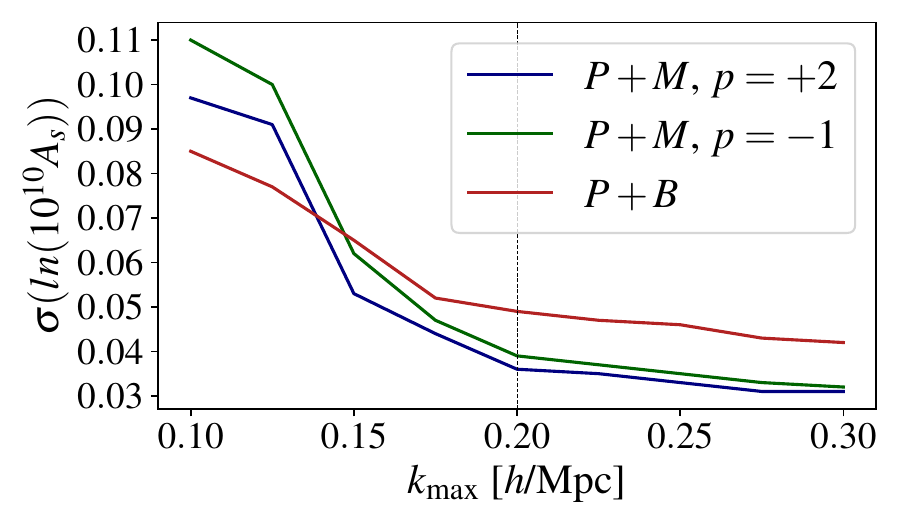}
    \includegraphics[width=0.325\linewidth]{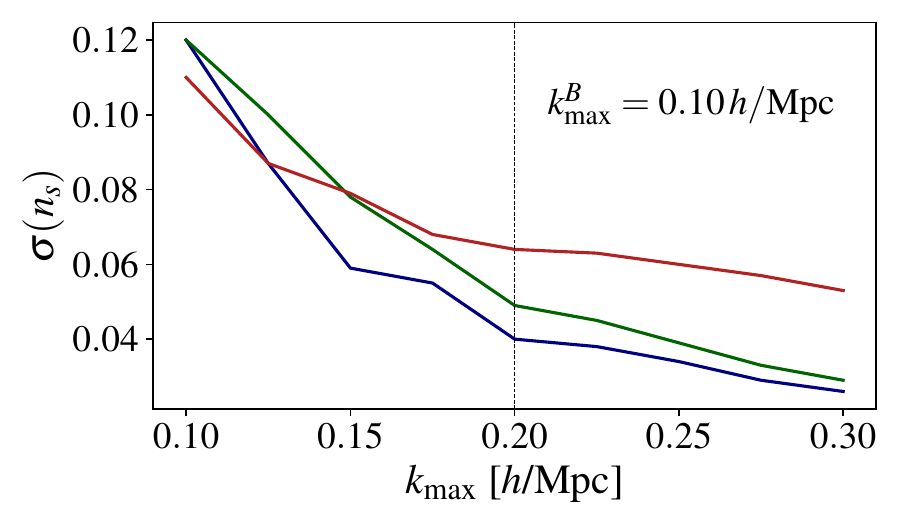}
    \includegraphics[width=0.325\linewidth]{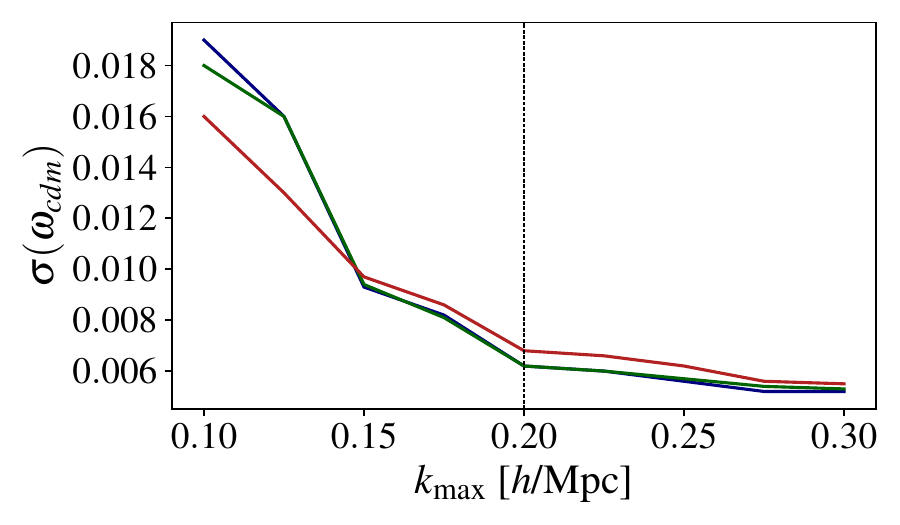}
    \includegraphics[width=0.325\linewidth]{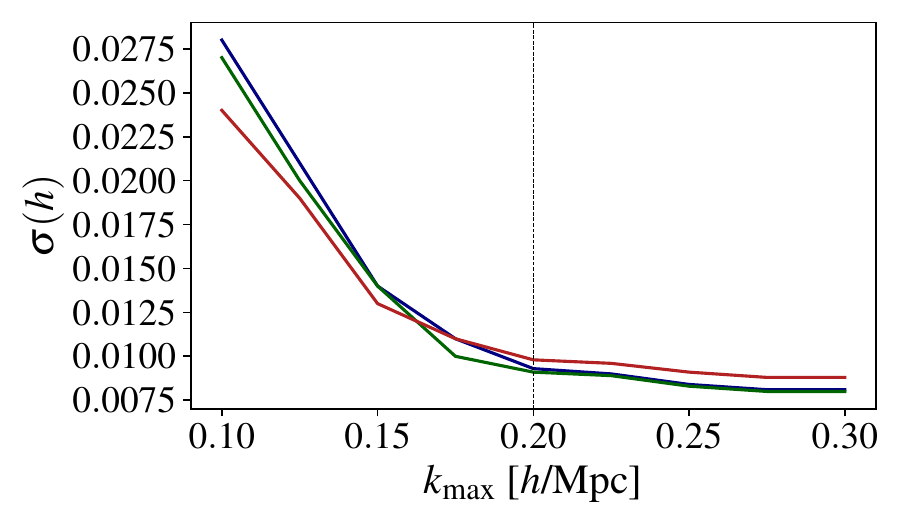}
    \includegraphics[width=0.325\linewidth]{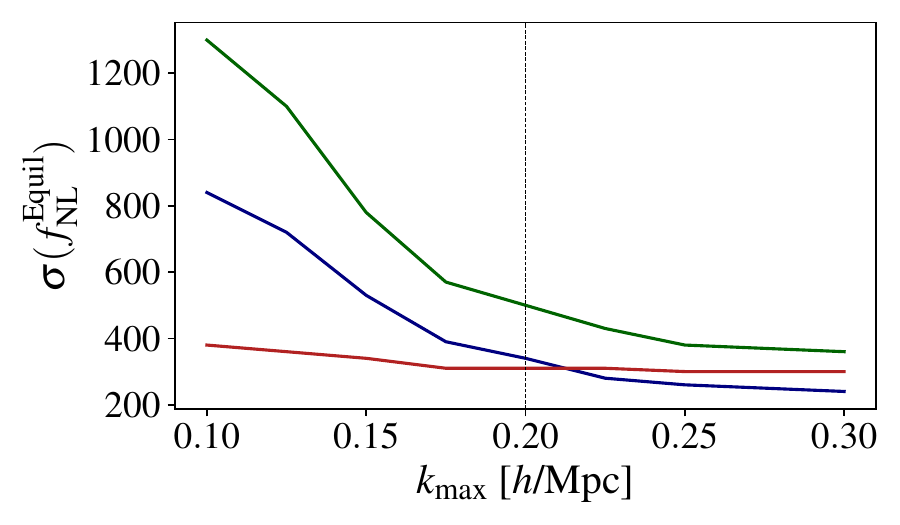}
    \includegraphics[width=0.325\linewidth]{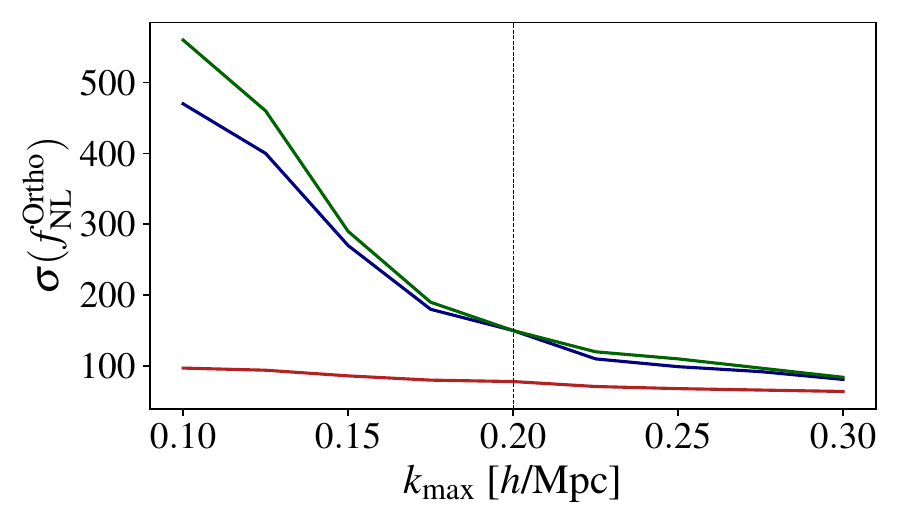}
    \caption{Same as figure~\ref{fig:matter_Mnu} but for cosmologies with primordial non-gaussianities.}
    \label{fig:matter_fnl}
\end{figure}

LSS observation always involve luminous sources (galaxies, quasars, $\dots$), hence studying the dark matter field is a merely academic exercise; nevertheless, it can still be worthy as it can give useful insights on where the physical information is. It also represents an optimal case, since the unknown bias parameters will degrade the constraints and the discrete nature of galaxies will diminish the signal-to-noise ratio due to the shot-noise. 
The results of our analysis are shown in fig.~\ref{fig:matter_Mnu} for the neutrino mass $M_\nu$ and fig.~\ref{fig:matter_fnl} for non-local PNG. We decide not to show the comparison between the standard power spectrum and the marked one as it has been shown that, with the same volumes and sky-cuts, the latter gives better constraints on almost all the cosmological parameters. 
we compare here two different mark models to understand if the information is coming from different structures of LSS web (voids as underdense regions or filaments and nodes for overdense ones) or it is just information taken form higher order correlation functions. The effects of massive neutrinos on LSS are mainly a rescaling of the growth of linear perturbation and a erasing of small scale clustering below the free streaming scale~\cite{Lesgourgues:2006nd}. The former is the most visible and detectable, while the latter is slightly degenerate with the small scales parameter, the counterterm $c_s^2$ that already appears at level of matter, and the higher order biases that are present for biased tracers.
For all the parameters, in the analysis where the total neutrino mass is varied, we can see that different marks, $p=2$ for voids and $p=-1$ for nodes, give constraints that are almost comparable, with slightly better results for positive $p$. Most importantly, we can see that these constraints are comparable with  the $P+B$ analysis, which, we remind, has a fixed $k_{\rm max}^{B} = 0.10\, h/\text{Mpc}$: this confirms that the additional information of the marked power spectrum is mostly a bispectrum-like contribution. In particular, for the neutrino mass $M_\nu$ we can see that for $k_{\rm max}>0.175 \,h/\text{Mpc}$, the constraints obtained are basically the same.

For the primordial non-gaussianities the outcomes are slightly different. Non-local non-gaussianities affect the standard power spectrum only through one-loop contribution (differently from the well-know large scales effect for non-gaussianities of the local type~\cite{Matarrese:2008nc, Desjacques:2008vf}) and hence their effect is either tiny and degenerate with higher order biases or counterterms. This means that the bispectrum plays a crucial role in constraining this kind of non-gaussianities as it is the main source of signal.
Figure~\ref{fig:matter_fnl} shows that the errors $\sigma(f_{\rm NL}^{\rm Equil}), \sigma(f_{\rm NL}^{\rm Ortho})$ from the $P+M$ analysis are always bigger than the $P+B$ one, at least on the scales at which EFT of LSS is still reliable for BOSS volumes. We see that the differences between a positive and a negative $p$ are negligible and that the results we obtain are very similar to those obtained with the bispectrum: this again confirms that the information content of the marked power spectrum is in fact a subset of the information contained in higher order correlation functions. While for the total neutrino mass this was not obvious as the contribution of the bispectrum to its constraints are not essential, in the case of non-local primordial non-gaussianities this results to be evident. 

At this stage, it is worth comparing our findings with the results of~\cite{Massara:2020pli}, where the authors investigated the constraining power of the marked power spectrum on massive neutrinos using a simulation based approach. Using the $cb$ (cold dark matter + baryons) component they find that using the marked power spectrum produces an improvement of $\sim 3$ over the standard one at $k_{\rm max} = 0.5\,h/\text{Mpc}$. If consider a direct comparison between $M$ and $P$, we find a similar improvement in our Fisher forecast -- at scales ($k_{\rm max}=0.20-0.30 \,h/\text{Mpc}$ -- when we consider the same volume, redshift and mark parameters. This again confirms the agreement with previous results, even though here we focus on the comparison with the standard $P+B$ analysis, which seems a more natural one, considering that $M$ does incorporate higher-order information.

The previous analysis in~\cite{Massara:2020pli} also emphasizes a void-enhancing mark is the optimal choice for constraining neutrino masses. This result is again qualitatively consistent with our findings for the matter field analysis, see figure~\ref{fig:matter_Mnu}. Moreover, we find that a similar conclusion holds for equilateral PNG's: as pointed out in previous works, see e.g.~\cite{Lewis:2011au}, the signal from equilateral PNG is maximized for non-linear filamentary structure and this could explain why a "void-enhanced" marked power spectrum could be effective in constraining this shape.

\subsection{Galaxies}

\begin{figure}[h]
    \centering
    \includegraphics[width=0.325\linewidth]{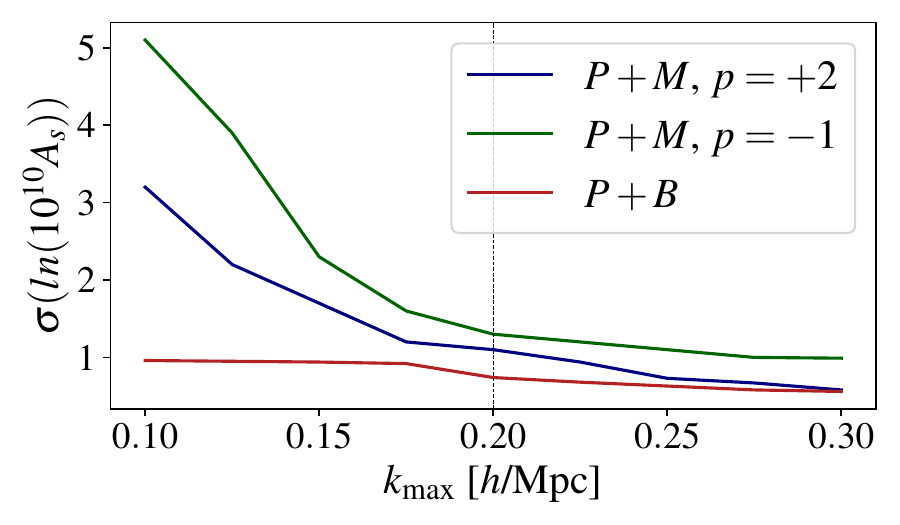}
    \includegraphics[width=0.325\linewidth]{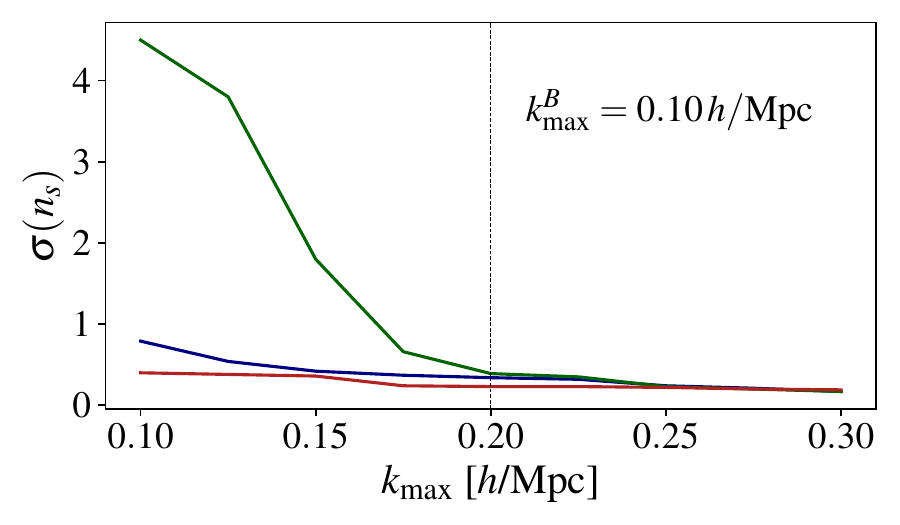}
    \includegraphics[width=0.325\linewidth]{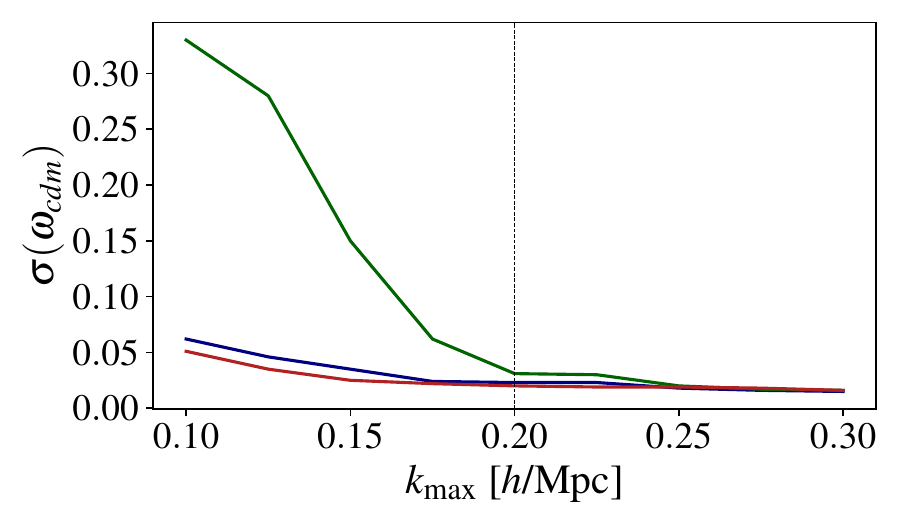}
    \includegraphics[width=0.325\linewidth]{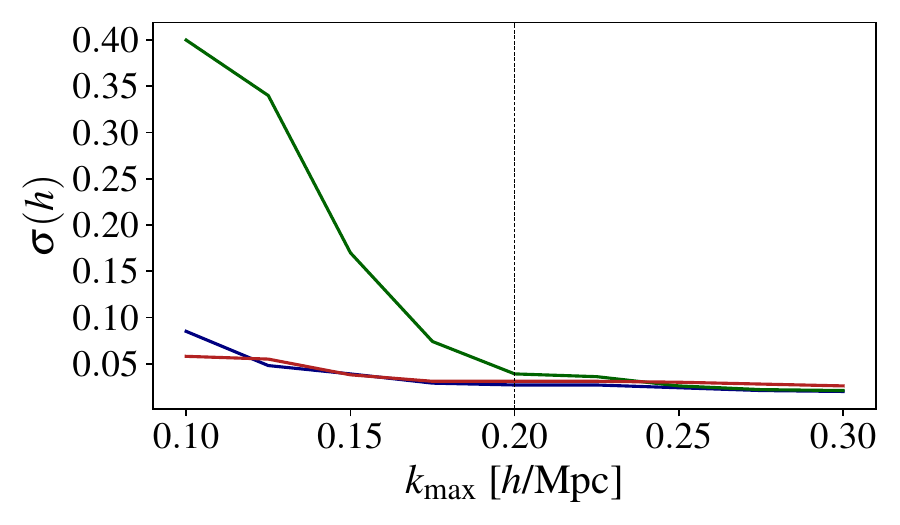}
    \includegraphics[width=0.325\linewidth]{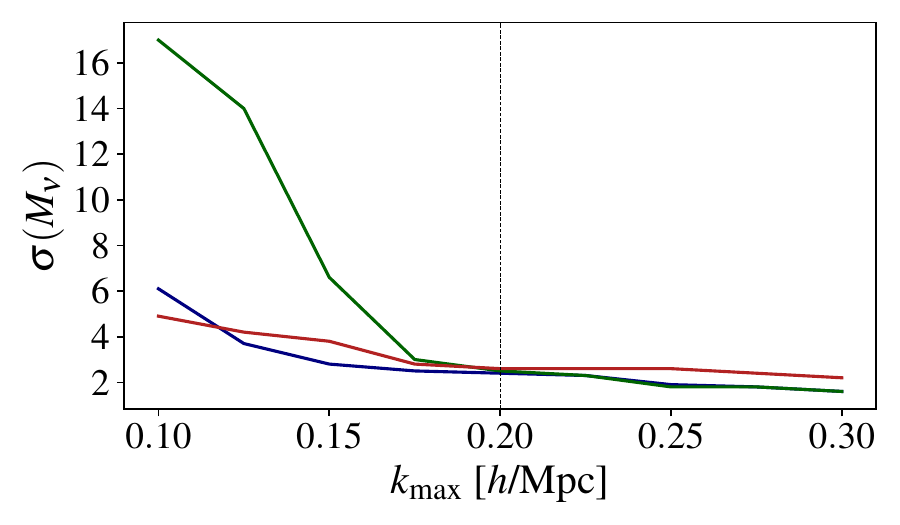}
    \includegraphics[width=0.325\linewidth]{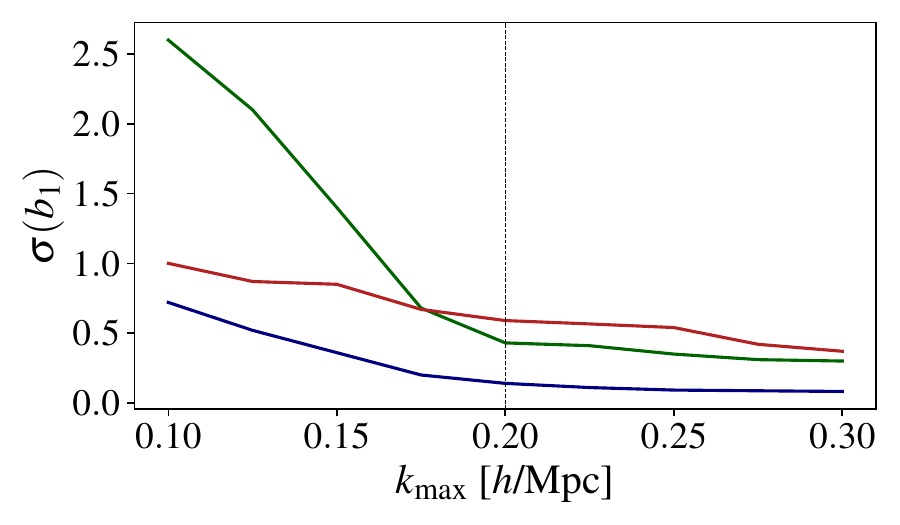}
    \includegraphics[width=0.325\linewidth]{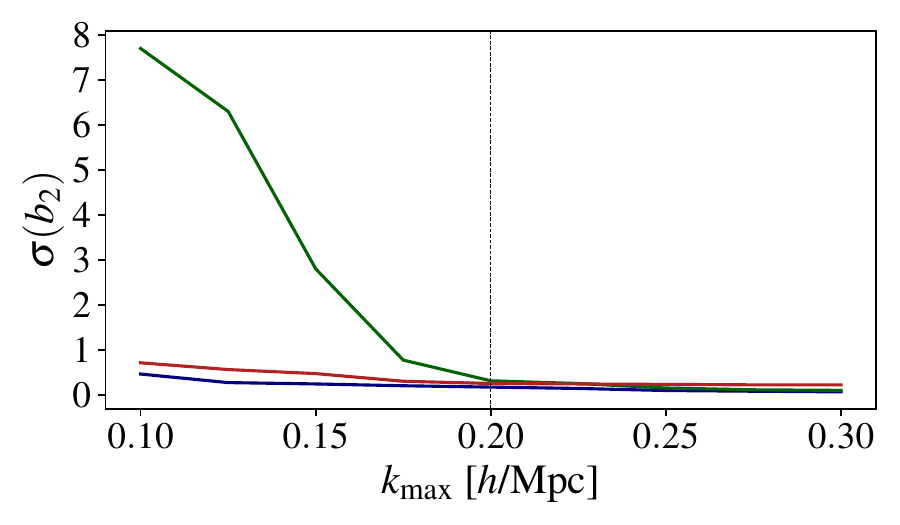}
    \includegraphics[width=0.325\linewidth]{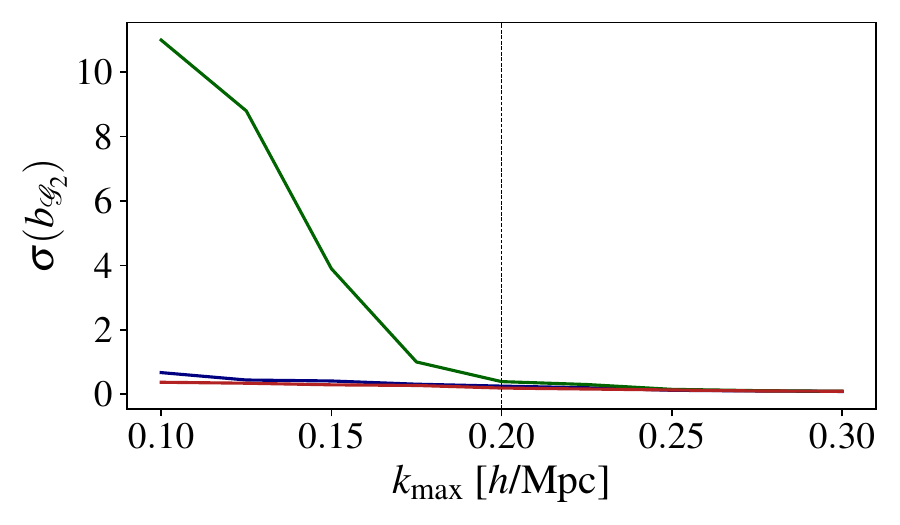}
    \caption{2$\sigma$ constraints from galaxies for massive neutrinos cosmologies and BOSS CMASS volume, number density and redshift. The dashed vertical line represent the maximum wave vector adopted in standard analysis~\cite{Ivanov_PS}.}
    \label{fig:gal_Mnu}
\end{figure}

Biased tracers, such as galaxies, introduce a certain degree of freedom through the biases, that parametrize our incomplete knowledge about the connection between dark matter and galaxies. They are the main source of degradation of cosmological information and indeed much effort has been recently made to constrain them using hydro-dynamical simulations~\cite{Ivanov:2024hgq, Ivanov:2024xgb}.
For what concerns our work, we know that, in real space, specific biases are almost perfectly degenerate with cosmological parameters. When the power spectrum alone is considered, the linear bias is completely degenerate with the primordial amplitude $A_s$, and this degeneracy can be slightly broken by using the information from the IR-resummed BAO. This degeneracy can be broken by combining the power spectrum and the bispectrum, at the price of introducing a new degeneracy among the scalar amplitude and the second order bias $b_2$. Including modes from smaller scales could indeed alleviate these degeneracies, allowing to constrain higher order biases through the one-loop terms. This eventually results in worse signal-to-noise for the cosmological parameters. For massive neutrinos we observe in figure~\ref{fig:gal_Mnu} that the marked power spectrum with $p=-1$ performs very bad for small $k_{\rm max}$'s, but it reach the level of constraints of $p=2$ when the loop terms are more important around $k_{\rm max}\simeq 0.2 \,h/\text{Mpc}$. 

Interestingly, we see that $\sigma(M_\nu)$ for the $P+M$ case, for maximum $k$ around the non-linear scale, is slightly better than the $P+B$ one, suggesting that we are taking more information from non-linear scales compared to the bispectrum at $k_{\rm max}^B = 0.10 \,h/\text{Mpc}$. It is also interesting to notice that the main effect of adding the bispectrum is usually a better measurement of the second order biases $b_2$ and $b_{{\mathcal G}_2}$, which are usually less constrained from $P$ alone analyses. We see that the inclusion of the marked power spectrum allows to measure these biases at the same level of $P+B$ analyses.

Neutrino mass affects the linear power spectrum through a modification of the linear growth and a damping around and below the free streaming scale, which is a nonlinear effect. The galaxy power spectrum alone is mostly sensitive to the former effects, while the latter is degenerate with higher order biases and counterterms. As shown in figure~\ref{fig:tri_bias_fnl}, the marked power spectrum, especially with the mark with $p=2$, the breaks some of the degeneracies among the total neutrino mass $M_\nu$ and the biased tracers, improving the constraints on the bias parameters. This should not change dramatically when we will consider redshift space distortions since, at first order, these will allow to have a better measurement of the linear bias $b_1$, which does not appear to be degenerate with the neutrino total mass. 

\begin{figure}[h]
    \centering
    \includegraphics[width=0.85\linewidth]{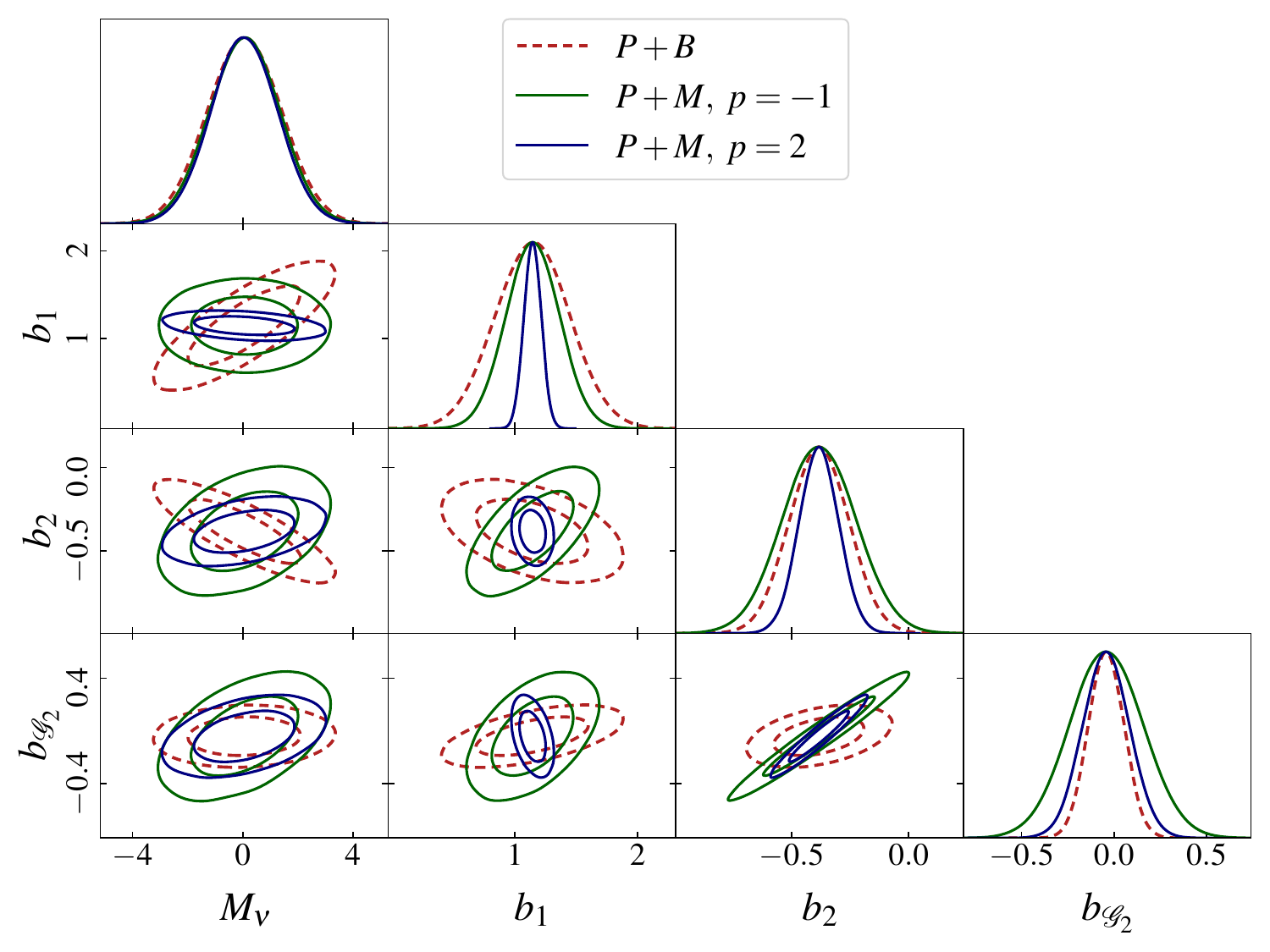}
    \caption{Triangle plot for massive neutrinos cosmology and biased tracers. We compare $P+M$ with $k_{\rm max}^{P,M} = 0.20 \,h/\text{Mpc}$ and $P+B$ with $k_{\rm max}^B = 0.10 \,h/\text{Mpc}$.}
    \label{fig:tri_bias_Mnu}
\end{figure}

\begin{figure}[h]
    \centering
    \includegraphics[width=0.325\linewidth]{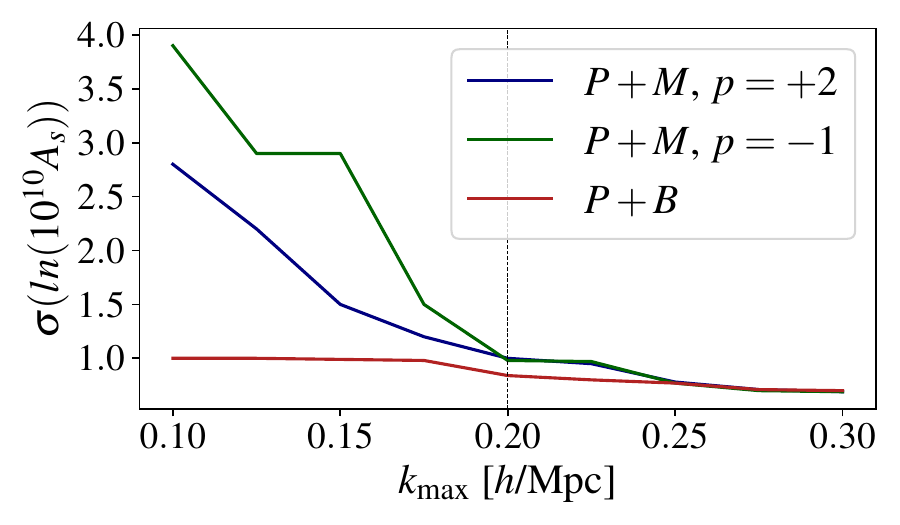}
    \includegraphics[width=0.325\linewidth]{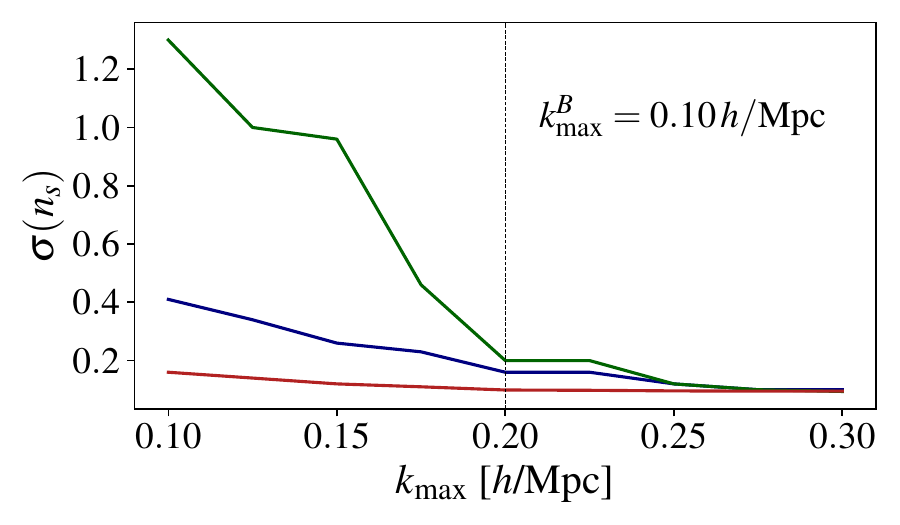}
    \includegraphics[width=0.325\linewidth]{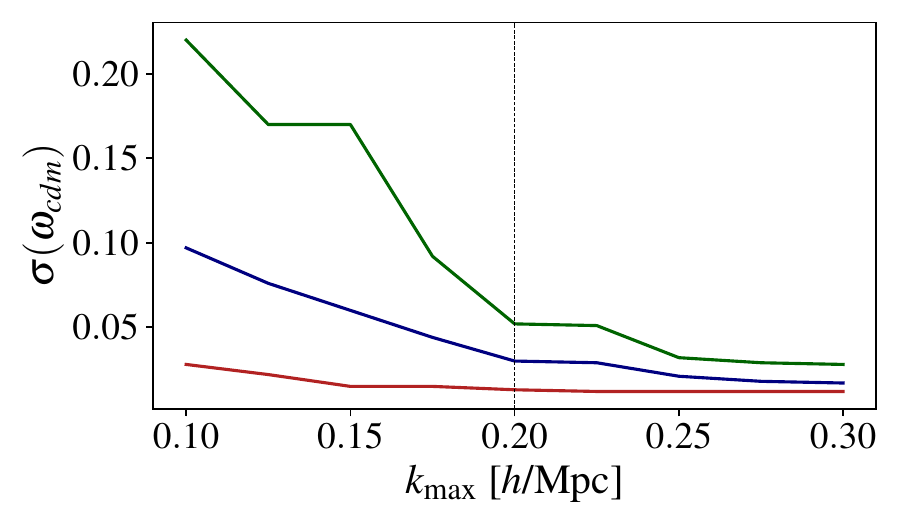}
    \includegraphics[width=0.325\linewidth]{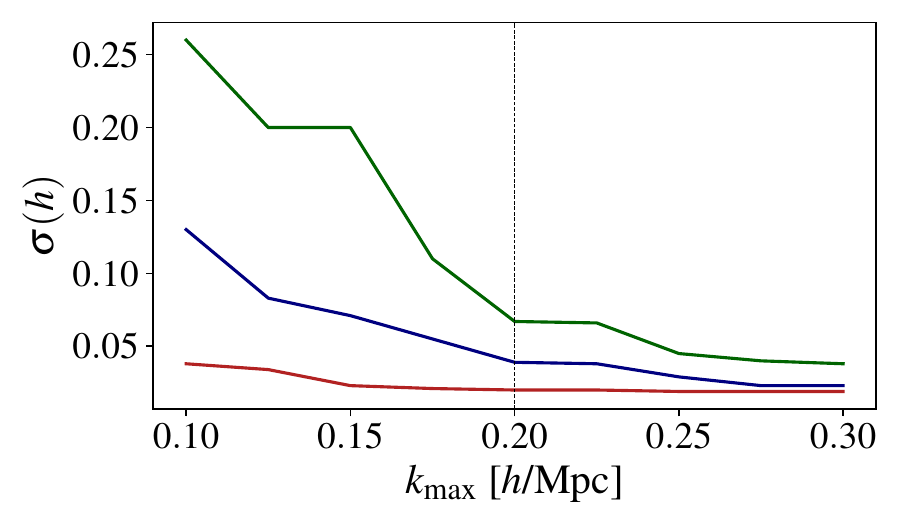}
    \includegraphics[width=0.325\linewidth]{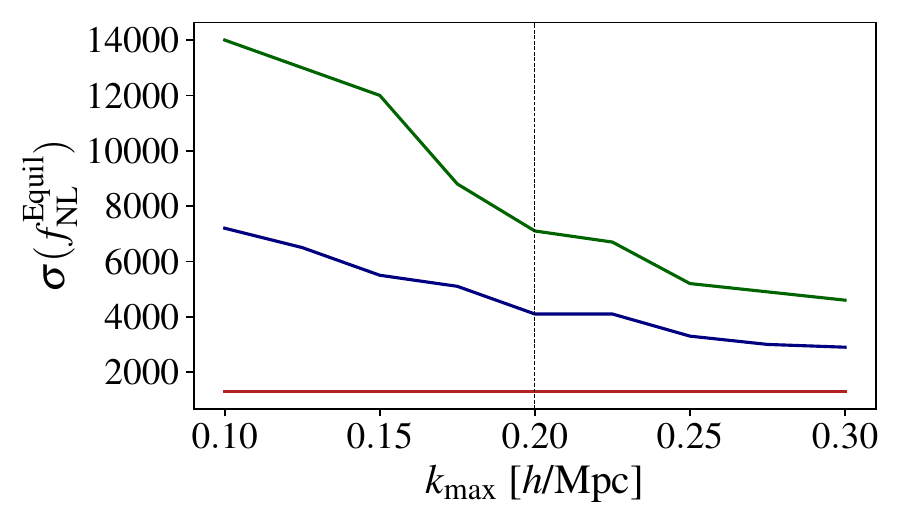}
    \includegraphics[width=0.325\linewidth]{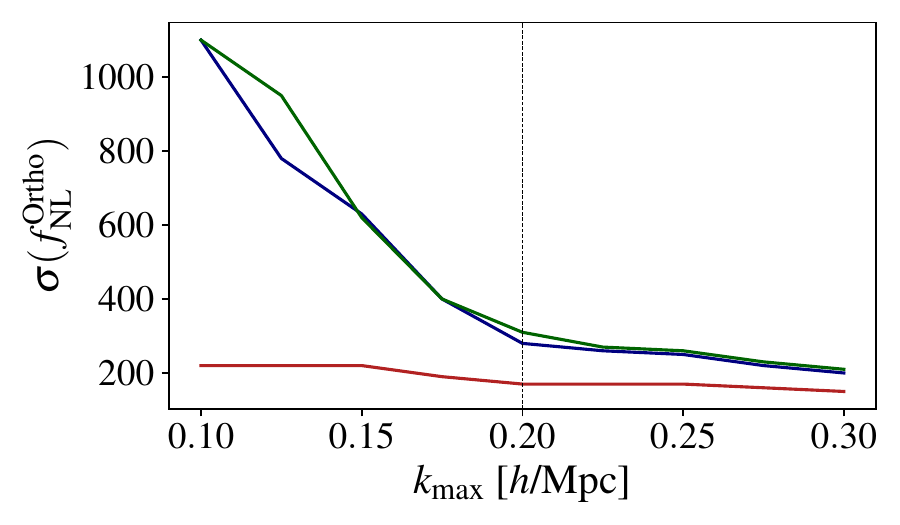}
    \includegraphics[width=0.325\linewidth]{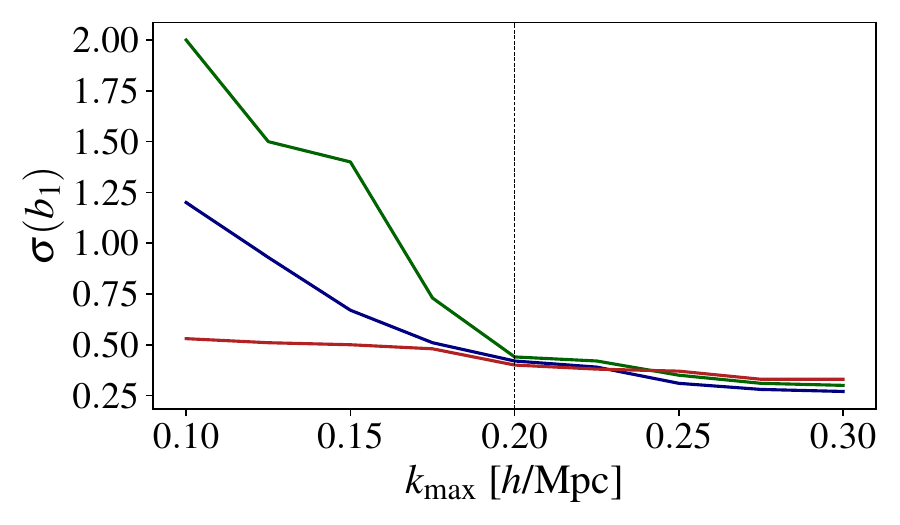}
    \includegraphics[width=0.325\linewidth]{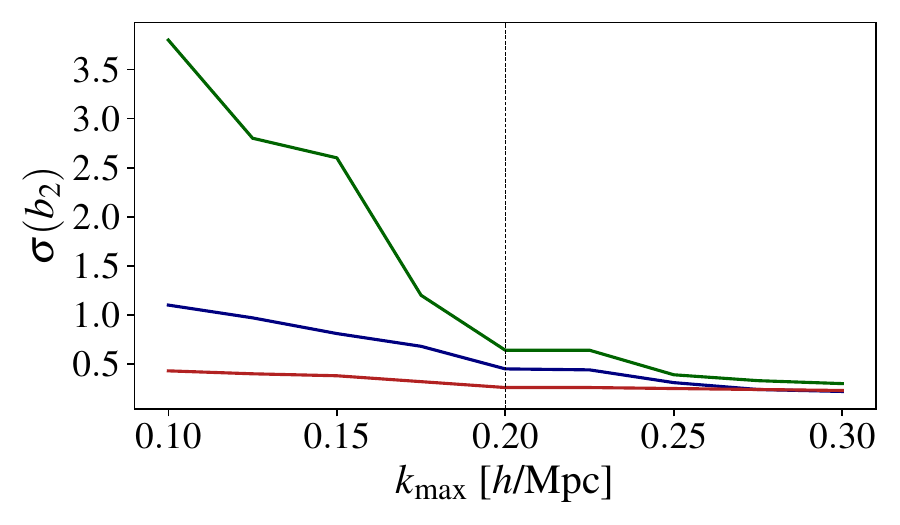}
    \includegraphics[width=0.325\linewidth]{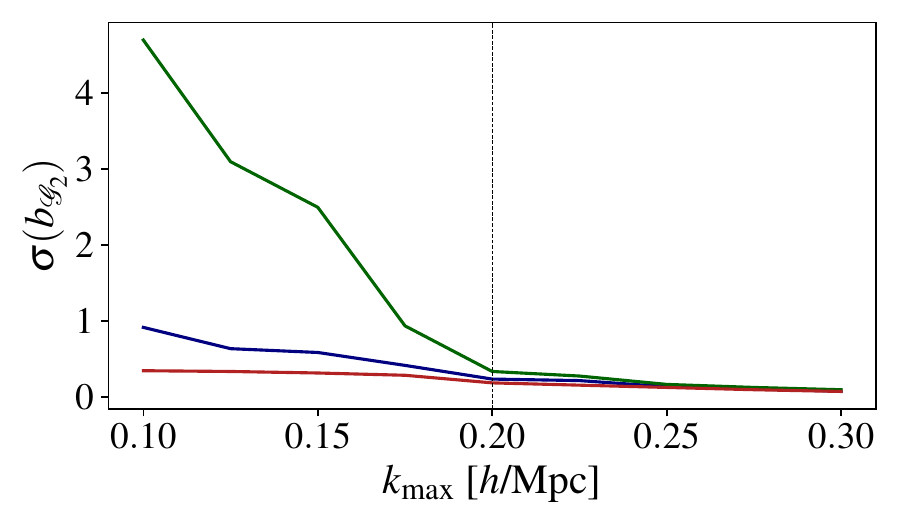}
    \caption{Same as figure~\ref{fig:gal_Mnu} for cosmologies with primordial non-gaussianities.}
    \label{fig:gal_fnl}
\end{figure}

For non-local primordial non-gaussianities, we see a similar behavior to what we found with the dark matter field. We already mentioned that for the shapes constrain here, the bispectrum is essential in order to constrain primordial non-gaussianities, differently from the case of massive cosmologies, where most of the information is taken from linear scales. The results of our analysis are shown in figure~\ref{fig:gal_fnl}. We notice that in both cases where $p=-1$ and $p=2$, the constraints on $f_{\rm NL}^{\rm Equil}$ and $f_{\rm NL}^{\rm Ortho}$ obtained combining $P$ and $M$ never reach those obtained with $P+B$, even for very non-linear $k_{\rm max}$ and the same is true for almost all the other parameters varied in this analysis. This is again a confirmation of what we mentioned earlier for massive neutrinos: if the information about a cosmological parameter within the galaxy power spectrum comes both from linear and non-linear effects, then the marked power spectrum can improve the constraints obtained when combining with the bispectrum, as it is for $M_\nu$\footnote{We stress again that usual analyses as in ref.~\cite{Jung_Quij} always compare $P$, $M$, $B$ and various combinations of them at the same $k_{\rm max} = 0.5 \,h/\text{Mpc}$. Here we are using the maximum scale at which the EFTofLSS is valid, since going to higher $k_{\rm max}$ would give biased results in realistic analyses due to exclusion of higher order loops entering in the analytical model. We are, however, in agreement with their findings.}; on the other hand, non-local PNG affect the galaxy power spectrum only through a one-loop effect, and basically all the information is contained in the bispectrum, hence the marked power spectrum can only do as well as the bispectrum does, for the scales considered.
As also pointed out in~\cite{Massara:2022zrf, Massara:2024cvu}, the degrading of the constraints on the neutrino masses is probably caused by the low density of the survey considered. This is again in agreement with our findings, shown in figure~\ref{fig:gal_Mnu}, which however show that for $k_{\rm max}\gtrsim 0.2 \text{Mpc}$ the $P+M$ works slightly better than the $P+B$ analysis. It is important however to stress that our results are obtained in real space and without accounting for redshift space distortions, survey mask effects, binning, which are expected to degrade even more our results.

For biased tracers we also see, as shown in figure~\ref{fig:tri_bias_fnl}, a small degeneracy among the PNG parameters and the linear bias $b_1$: this is a consequence of the universality relation assumed for the non-Gaussian bias in eq.~\ref{eq:bzeta}.

\begin{figure}[h]
    \centering
    \includegraphics[width=0.85\linewidth]{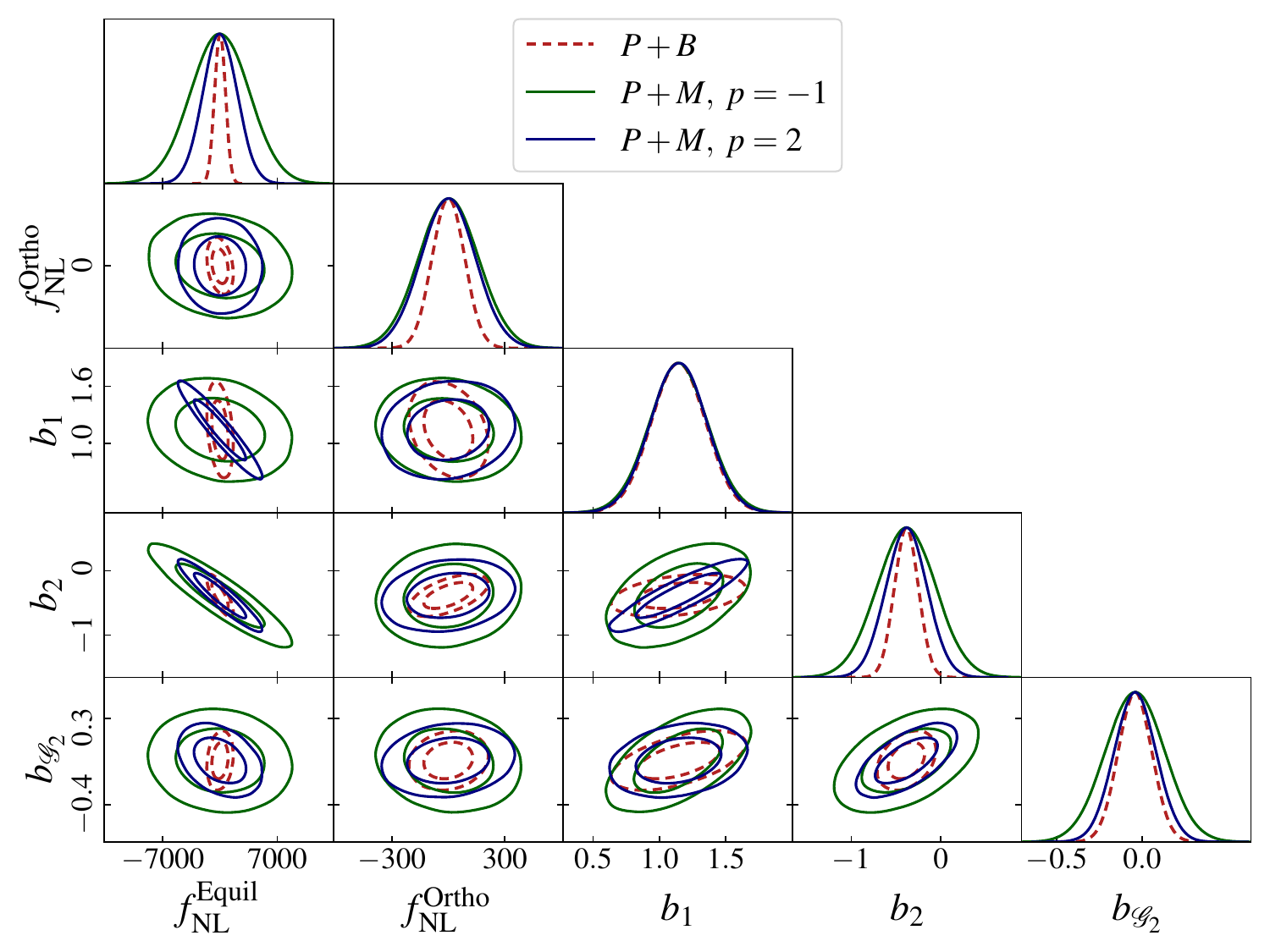}
    \caption{Same as figure~\ref{fig:tri_bias_Mnu} but for primordial non-gaussianities.}
    \label{fig:tri_bias_fnl}
\end{figure}

The corner plots presented in this work suggest that combining the power spectrum and the bispectrum with different marks would give even better results either for massive neutrinos and primordial non-gaussianities. This kind of combination is beyond the scope of this work and we will investigate it in future work accounting also for redshift space distortions and other observational effects.

\section{Discussion and conclusion}

In this work we present the first analytical study of the marked power spectrum in the context of primordial non-Gaussianities (PNG) of the non-local type, contributing new insights into the behavior and potential applications of marked statistics in cosmology. Following the work of~\cite{Philcox_1, Philcox_2} we develop a theoretically consistent model for the marked power spectrum at one-loop order adopting the Effective Field Theory of Large Scale Structure to reach mildly non-linear scales and analytically establish the information content of such statistic. When compared to the Quijote-PNG simulations~\cite{Coulton:2022qbc}, our quantitative results reveal a strong agreement between analytical predictions and simulations, supporting the robustness of the theoretical method adopted to describe the marked power spectrum. The model displays an excellent agreement with simulated data from low redshifts ($z=0$) up to higher ones ($z=1$), suggesting that it can be reliably adopted also for on-going and future large volume surveys, such as the DESI~\cite{DESI:2016fyo} and Euclid\cite{Cimatti:2009is, Euclid:2024yrr} missions. The power of marked power spectrum resides in the possibility to enhance some features of the non-linear galaxy distribution, in particular by up-weighting underdense regions such as voids, where some cosmological information is expected to be found. 

In this work, we compare the constraining power on non-local PNG and massive neutrinos -- obtained through a combined analysis of power spectrum and marked power spectrum, $P+M$ -- with the traditional and well-established power spectrum and bispectrum, $P+B$, approach. We find that marking underdense regions produces the highest signal, in terms of error bars on $M_\nu$ and both $f_{\rm NL}^{\rm eq}$ and $f_{\rm NL}^{\rm or}$, in comparison to up-weighting higher density regions. Neither choice however improves over the $P+B$ method, within the mildly non-linear scale range investigated in this work. More precisely, we find that up-weighting voids is efficient in a low shot-noise context, as it saturates the $P+B$ signal in our pure dark matter field analysis. On the other hand, when considering realistic, discrete, biased tracers with BOSS-like galaxy number density and volume, the gain is limited and $P+B$ turns out to be the most optimal approach. This is in agreement with previous forecasts~\cite{Massara:2020pli, Massara:2022zrf} and with the recent data analysis performed on the BOSS data using a simulation based approach~\cite{Massara:2024cvu}. These results are however expected to improve with future low-shot-noise surveys, where underdense regions would play an important role.

Even though the $P+M$ combination does not produce improvements over $P+B$, the marked power spectrum is a promising tool to constrain beyond-$\Lambda$CDM physics, such as PNG, massive neutrinos, and modified gravity~\cite{Karcher:2024twr}. A $P+B$ analysis based on the EFT approach still requires the implementation of an estimator for the bispectrum and to take into account a number of complex observational effects, such as binning and convolution with the survey window. Since the marked power spectrum is a two-point statistic, these effects have a less severe and well understood impact, which is much easier to keep under control. Moreover, it was recently shown that not including off-diagonal terms in the bispectrum covariance induces an underestimation of the error bars for cosmological parameters, especially for primordial non-gaussianities~\cite{Floss:2022wkq}, which are the focus of this work. The diagonal approximation for the marked power spectra covariance is, on the other hand, more accurate, as tested in~\cite{Massara:2020pli, Massara:2022zrf}. This makes the marked power spectrum potentially more powerful at extracting NG information at non-linear scales; at the same time, this is the reason why we adopt a Gaussian approximation for the marked power spectrum covariance in our forecasts. 

In the present work we adopt a perturbative model whose validity is truncated to mildly non-linear scales $k_{\rm max} \simeq 0.25\, h/\text{Mpc}$, while simulation-based approaches~\cite{Massara:2024cvu, Jung_Quij} could potentially reach very non-linear scales $k_{\rm max} = 0.50 \,h/\text{Mpc}$ where more information from the cosmic web could be captured. This is relevant for neutrino mass detection, modified gravity and PNG, where the small-scale, non-linear behavior offers significant leverage for parameter constraints.

Our findings also align with the recent work~\cite{2024arXiv240917133E}, where the authors explore new expressions for the mark function and investigate the degeneracy-breaking power of the marked power spectrum, compared to a power spectrum only analysis. In this work, we further extend this by finding, indeed, that this constraining power on higher order biases, such as $b_2$, of the marked power spectrum, is indeed descending from the inclusion of bispectrum-like convolution terms, that help in breaking some of the crucial degeneracies. This comparison reinforces the value of continued development in this analytical framework and further highlights the marked approach as a promising complement to established methodologies.

In future research, it will be natural to extend the present approach to redshift space, even if this will most likely not modify the core findings of this work about the amount of information gained with marked statistics. We will also need to include all the observational effects, such as Alcock-Paczynski, binning, and survey geometry, in order to build a full pipeline to use in actual data analysis, to further examine the capabilities and limitations of marked statistics in cosmological contexts. 

Another important direction is the optimization of the mark function, as it has been performed in~\cite{2024arXiv240905695C} for $\sigma_8$ and $\Omega_m$. Regarding the mark itself, we limited our exploration to two values for the parameter $p$, in order to up-weight either voids or nodes in the cosmic structure, using a fixed expression for the mark, eq.~\ref{eq:mark_here}. Our analytical approach -- possibly combined with numerical techniques (such as the Gaussian process-based optimization in~\cite{2024arXiv240905695C}) -- should lead to a generalization of the choice of the mark function, able to maximize PNG information. Such optimization could significantly boost the utility of marked statistics for addressing key questions in cosmology and will be investigated more in future works.

In summary, our study contributes to provide a foundational framework for the application of marked statistics in cosmological analyses -- with emphasis on PNG studies -- demonstrating their compatibility with simulation data, potential for parameter estimation, and promise for further refinement. By offering a distinct perspective from traditional $P+B$ analysis, marked statistics expand the cosmologist’s toolkit, opening new paths toward probing the properties of the primordial perturbations and the detailed composition of the universe.

\section*{Acknowledgments}
We are grateful to Thomas Fl\"oss, Elena Massara, Azadeh Moradinezhad, Massimo Pietroni and Marco Peloso for useful discussions during the development of this work. MM is particularly grateful to Marko Simonović for his thorough support with the FFTLog. MM and ML acknowledge support by the MUR Progetti di Ricerca di Rilevante Interesse Nazionale (PRIN) Bando 2022 - grant 20228RMX4A, funded by the European Union - Next generation EU, Mission 4, Component 1, CUP C53D23000940006. GJ acknowledges support from the ANR LOCALIZATION project, grant ANR-21-CE31-0019 / 490702358 of the French Agence Nationale de la Recherche. The Flatiron Institute is supported by the Simons Foundation.

\bibliographystyle{JHEP}
\bibliography{main}
\end{document}